\documentclass[letterpaper,twocolumn,10pt]{article}
\usepackage{usenix}

\usepackage[
 toc,              
 acronym,          
 section=chapter,  
 nonumberlist,     
]{glossaries}
\usepackage{xcolor}
\usepackage{graphicx}
\usepackage{subcaption}
\usepackage{booktabs}
\usepackage{tabularx}
\usepackage{xurl}
\usepackage{algorithm}
\usepackage{algorithmic}
\usepackage{pifont}
\usepackage[dvipsnames]{xcolor}

\newacronym{ipfs}{IPFS}{InterPlanetary File System}
\newacronym{dns}{DNS}{Domain Name System}
\newacronym{url}{URL}{Uniform Resource Locator}
\newacronym{cid}{CID}{Content Identifier}
\newacronym{pid}{PID}{Peer Identifier}
\newacronym{dht}{DHT}{Distributed Hash Table}
\newacronym{tls}{TLS}{Transport Layer Security}
\newacronym{dag}{DAG}{Directed Acyclic Graph}
\newacronym{ip}{IP}{Internet Protocol}
\newacronym{bgp}{BGP}{Border Gate Protocol}
\newacronym{rpki}{RPKI}{Resource Public Key Infrastructure}
\newacronym{scion}{SCION}{Scalability, Control, and Isolation On Next-Generation Networks}
\newacronym[plural=ASes, firstplural=ASes]{as}{AS}{Autonomous System}
\newacronym{ixp}{IXP}{Internet Exchange Point}
\newacronym{rpc}{RPC}{Remote Procedure Call}
\newacronym{dos}{DoS}{Denial of Service}
\newacronym{api}{API}{Application Programming Interface}

\begin{document}

\title{Network-level Censorship Attacks in the InterPlanetary File System}


\author{
{\rm Jan Matter}\\
ETH Zürich \\
jan.matter@outlook.com
\and
{\rm Muoi Tran}\\
Chalmers University of Technology and University of Gothenburg \\
muoi@chalmers.se
} 



\maketitle

\begin{abstract}
The InterPlanetary File System (IPFS) has been successfully established as the de facto standard for decentralized data storage in the emerging Web3. 
Despite its decentralized nature, IPFS nodes, as well as IPFS content providers, have converged to centralization in large public clouds \cite{balduf2023cloud} \cite{cheng2025centralization}. 
Centralization introduces BGP routing-based attacks, such as passive interception and BGP hijacking, as potential threats. 
Although this attack vector has been investigated for many other Web3 protocols, such as Bitcoin and Ethereum, to the best of our knowledge, it has not been analyzed for the IPFS network. 
In our work, we bridge this gap and demonstrate that BGP routing attacks can be effectively leveraged to censor content in IPFS. 
For the analysis, we collected 3 $\times$ 1000 content blocks called CIDs and conducted a simulation of BGP hijacking and passive interception against them. 
We find that a single malicious AS can censor 75\% of the IPFS content for more than 57\% of all requester nodes. 
Furthermore, we show that even with a small set of only 62 hijacked prefixes, 70\% of the full attack effectiveness can already be reached. 
We further propose and validate countermeasures based on global collaborative content replication among all nodes in the IPFS network, together with additional robust backup content provider nodes that are well-hardened against BGP hijacking. We hope this work raises awareness about the threat BGP routing-based attacks pose to IPFS and triggers further efforts to harden the live IPFS network against them.
\end{abstract}
\section{Introduction}
\label{sec:introduction}

IPFS was introduced as an idea in 2014 by Juan Benet to create a distributed, uncensorable, and permissionless file system~\cite{benet2014ipfs}. 
In the following years, IPFS was implemented and evolved into an extensive peer-to-peer (P2P) network with numerous production use cases in the emerging Web3~\cite{ipfs_origins_2013}. 
The IPFS network was measured in 2022 to consist of at least 50k peers~\cite{daniel2022passively}. 
The main contribution of IPFS lies in a novel way to look up content based on immutable content-based addresses called CIDs, which are derived from splitting a file into smaller chunks and building a Merkle DAG over its hashes. 
This solution allows content to be stored and addressed, while every retrieving party can verify the integrity of the files. 
Furthermore, IPFS uses the Kademlia Distributed Hash Table (DHT)~\cite{maymounkov2002kademlia}, which enables efficient storage and retrieval of CID provider records. 
This is achieved by storing the provider records on the peers with the minimal XOR distance between the CID and the peer's peer ID. 
Using the DHT, any CID provider record can be efficiently searched in an iterative search in $\log(n)$ time. 
Thus, it is possible to provide immutable content on the IPFS network without relying on any centralized resolution mechanism, such as DNS providers, which historically were repeatedly targets of DoS attacks~\cite{sommese2022ddos,coldewey2020cloudflare,siddiqui2018route53}.

IPFS, having gained increasing popularity, has attracted the interest of researchers investigating the performance and security aspects of the live IPFS network. 
Passive monitoring~\cite{balduf2022monitoring,daniel2022passively} and active crawling ~\cite{henningsen2020mapping} have enabled a systematic understanding of different aspects of the IPFS network, including its topology, size, distribution, and performance. 
Furthermore, specific weaknesses with corresponding exploitation possibilities in the IPFS protocol were found. 
An inherent risk facing IPFS is the usage of Sybil nodes. 
Several possible attack vectors were found~\cite{prunster2020eclipse,sridhar2024censorship} and have been mitigated thereafter.

In our work, we analyze the possibility that an attacker can censor content in the form of CIDs on the IPFS network by making it unavailable to potentially requesting IPFS nodes. 
We model the attacker to have full control over an AS by controlling all its BGP announcements, as well as how inflowing and outflowing traffic is manipulated, dropped, or forwarded. 
This assumption aligns well with other similar works in the area of BGP-based cryptocurrency attacks~\cite{apostolaki2017hijacking, saad2023three, tran2024routing}. 
Furthermore, we assume that the adversary hosts a single IPFS node to collect relevant attack data on the IPFS network. 
These resources are fairly easy to acquire for a few thousand dollars a month. 
Especially for nation-state actors who are one of the most relevant censorship actors today, these costs are well within reach~\cite{raman2020worldwide}. 
As an example of a real-world censorship attempt that also affected IPFS, an incident during the Catalan referendum in 2017 can be presented. 
The Spanish government at the time censored a voting app that used IPFS to circumvent its blocking on the Play Store and other download sources. 
While the Spanish government did not attempt to fully block IPFS directly, they did block the domain for \url{gateway.ipfs.io}, which leads to an IPFS API gateway that allows web apps to access IPFS content without running their own node. This incident clearly shows that IPFS censorship is a real-world concern \cite{spain-block-ipfs}.

To the best of our knowledge, there is no existing publication investigating AS-level network attacks, such as passive traffic interception or BGP hijacking, against the IPFS network. 
At the same time, traffic interception attacks against other peer-to-peer networks, such as Bitcoin or Ethereum, have been regularly identified and mitigated~\cite{tran2024routing,fan2021conman,doumanidis2025routing}. 
In traffic interception attacks, the attacker controls an AS that naturally intercepts targeted traffic flows. 
Once the traffic reaches the attacking AS, manipulation, dropping, or eavesdropping on the packets potentially allows the attacker to undermine the confidentiality, integrity, or availability of the service that depends on the traffic. 
Since in today's BGP-based Internet, senders still cannot actively select routes, and route announcements are not fully verified even with RPKI enabled, interception attacks are still a severe security risk for any service that requires communication over the Internet. 
Although the decentralized nature of P2P networks can help mitigate these risks, centralized components are often still vulnerable to effective attacks. 
As we show in our work, IPFS is no exception in that regard.


A key role in interception attacks is the centralization of key components of the target system in as few IP prefixes as possible. 
IPFS aims to build around decentralization, which should shield it against interception attacks. 
However, as previously shown, the live IPFS network shows centralization of its nodes and content providers in a small set of cloud providers~\cite{balduf2023cloud, cheng2025centralization}. 
Furthermore, a key observation is that content is often only hosted on a few provider nodes, often sitting in a small set of ASes and IP prefixes. 
These observations were the basis for our attack, which aims to allow for the denial of access to content for specific requester nodes over the IPFS network.


\begin{figure}[t!]
\begin{center}
  \includegraphics[width=0.48\textwidth]{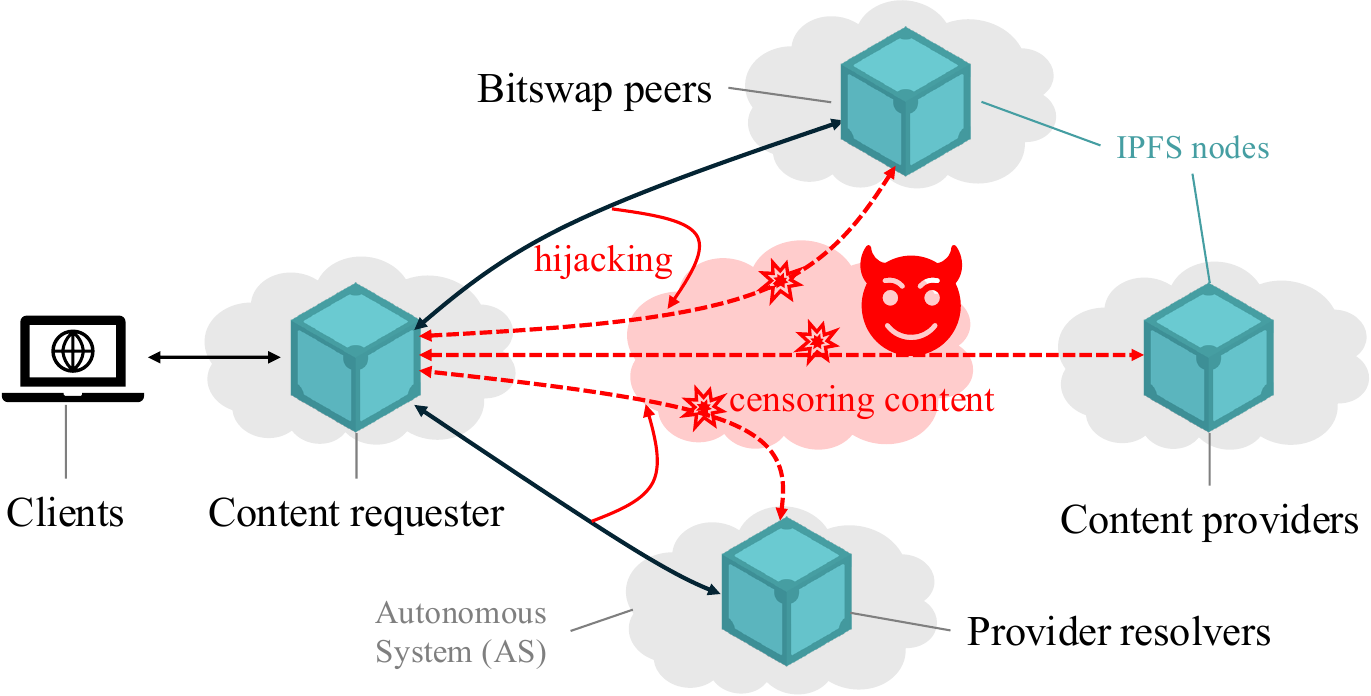}
  \caption{High-level overview of routing-based censorship in IPFS.
The malicious AS censors the requester's traffic by dropping it, either through passive interception or BGP hijacking, disrupting its exchanges with providers (including Bitswap peers), or blocking its provider discovery via resolvers.}
  \label{fig:attack-overview}
\end{center}
\end{figure}


To attack content retrieval on the IPFS network with an interception-based attack, we propose two possible attack vectors (see Figure~\ref{fig:attack-overview}). 
For both attack vectors, the attacker must block all connections to peers that cache the content and can share it directly through the Bitswap protocol. 
Furthermore, on the one hand, all connections to peers that hold provider records can be blocked, essentially denying the resolution of the provider IP addresses. 
Alternatively, the connections to all content providers can be blocked to block the target from receiving content for a specific CID from its providers, thus creating the second attack vector. 
A combination of blocking both the resolvers and the providers is expected to yield the maximum blockage rate.

Some of the key results of our work are that only a small percentage, between 0.5 and 8\% of all CIDs can be fully blocked from all requesters. 
However, for more than 75\% of the CIDs, a blockage is possible from more than 57\% of the requesters. 
A further fascinating observation is that a minimal number of only 62 hijacked IP prefixes is sufficient to achieve an attack effectiveness of more than 70\%, making the attack significantly more cost-effective compared to a full blockage, which would require the hijacking of several thousand IP prefixes. 
We were also able to demonstrate that using a global collaborative content provisioning protocol in IPFS can already reduce the expected blockage rate that an attacker could achieve to approximately 20\% of all requesters.

We claim the following contributions:

\begin{itemize}
  \item We investigate the centralization of content providers and content provider resolvers in the IPFS network based on their AS and IP prefix distribution.
  \item We simulate the effects of passive interception and BGP hijacking on the live IPFS topology. To evaluate if there is content that can be blocked for all requesters and to quantify, for content that cannot be entirely blocked, the fraction of requesters for which it can be blocked. Additionally, we investigate the number of IP prefixes that need to be blocked to effectively censor a large portion of content from most requesters.
  \item We suggest and simulate several mitigation techniques to protect content as a content provider and across the entire network, thus enhancing content protection in IPFS against traffic interception attacks.
\end{itemize}





\section{Background}
\label{sec:background}

In this section, we provide the context for network-level censorship attacks in IPFS. 
We start with a brief introduction to the IPFS network, focusing on nodes, peering connections, and some other relevant components (\S\ref{subsec:ipfs}).
We then describe the core operations of IPFS, including publishing and retrieving content (\S\ref{subsec:content}).
Furthermore, we provide a brief explanation of network-level attack capabilities, focusing on active BGP hijackings and passive traffic interception. (\S\ref{subsec:routing-attacks}).

\subsection{IPFS Network}
\label{subsec:ipfs}

IPFS operates on top of a P2P network of nodes implemented by the libp2p library \cite{libp2p}.
In addition, IPFS maintainers incorporate a few centralized components to address performance challenges when using IPFS as well as to allow its usage without having to run a full IPFS node, for example, for web clients.


\paragraph{Nodes.}
Each node on the IPFS network possesses a unique public–private key pair, from which a peer ID is derived by hashing the encoded public key into a multihash ~\cite{multihash}. By default, nodes keep their peer IDs unchanged over their lifetime.
To participate in the P2P network, the nodes advertise one or more multiaddresses, specifying the IP addresses, transport protocols, and port numbers associated with a given peer ID. A multiaddress is a simple data structure that allows a peer ID to be linked to multiple endpoints (e.g., for multihoming). Conversely, a single IP address may correspond to numerous peer IDs, e.g., when multiple nodes run on the same machine or when they are located behind a NAT.

\paragraph{Peers.}
IPFS employs a distributed Kademlia hash table (DHT) ~\cite{maymounkov2002kademlia} to maintain node records in the network. 
When a node first joins, it contacts one of several pre-configured bootstrap nodes, allowing the newcomer to query the DHT over the bootstrap node.
Random peer ID lookups allow the connecting IPFS node to find other peers in the network to which it can then connect.
Depending on whether the incoming IPFS node is connectable by other peers, it decides to set itself as an IPFS server or client node.
Only IPFS server nodes maintain a DHT routing table and respond to routing requests from other peers. 
The new node also performs a DHT walk towards its own peer ID, enabling verification that no other node in the network has the same ID, and allowing it to discover additional peers to populate its DHT routing table.
More specifically, the routing table contains buckets for each prefix length ranging from 0 to 255. Each bucket contains up to $k=20$ peer records, sharing in their peer ID the first N bits with the node's peer ID, where N is the position of the bucket. 
Connections that are made during the joining process typically remain open until they are cut to a configured range by the so-called ConnMngr. The number of open connections for a typical IPFS server node is generally in the range of a few hundred. 
Establishing a P2P connection involves negotiating a transport protocol, a secure channel protocol, and a multiplexing protocol, followed by creating a secure stream for encrypted, authenticated, and integrity-protected communication (e.g., with TLS 1.3)~\cite{libp2p-tls}.

\paragraph{Centralized components.}
Due to performance limitations in publishing and retrieving content (as shown later in \S\ref{subsec:content}), IPFS maintainers have introduced three centralized components: Interplanetary Network Indexers (IPNI), Hydra Boosters, and HTTP Gateways~\cite{wei2024eternal}. 
In particular, an IPNI is a high-performance server that maintains an index of popular content providers. 
By storing this mapping directly, it eliminates the need for the network to perform the usual lookup that converts content CIDs into provider peer IDs, which significantly reduces retrieval latency.
Hydra boosters are specialized IPFS nodes with multiple IPFS node heads, enabling them to maintain thousands of simultaneous peer connections. 
This high connectivity allows them to return routing information close to the target during DHT walks, thus significantly reducing the number of hops required to locate content.
Additionally, HTTP gateways are Nginx-based web servers that are bundled with IPFS nodes, bridging the traditional web and the IPFS network. 
They enable users to access IPFS content by issuing standard HTTP GET requests to a gateway URL, eliminating the need for the client to install and run an IPFS node.

\subsection{Publishing and Retrieving Content in IPFS}
\label{subsec:content}

\begin{figure*}[t!]
\begin{center}
  \includegraphics[width=0.8\textwidth]{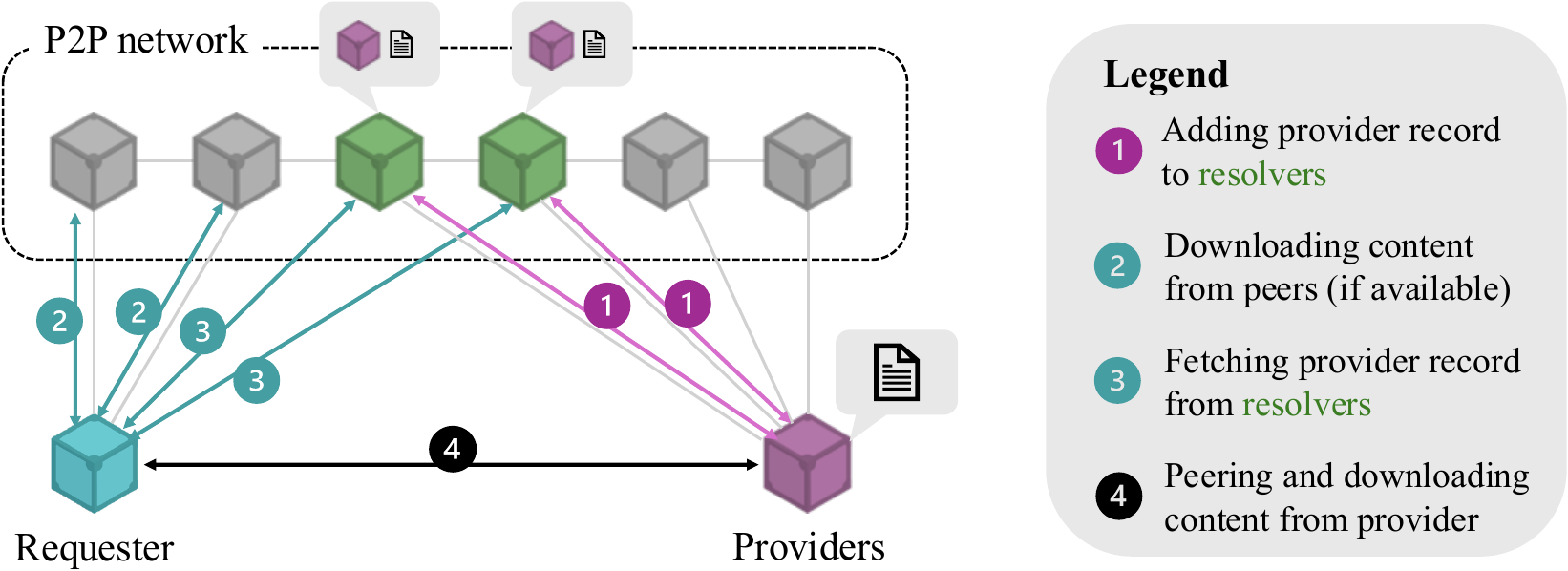}
  \caption{Publishing and Retrieving Content in IPFS.  The content provider binds the content $C$ and its peer ID into provider records and publishes them to $k$ peers closest to $C$, referred to as resolvers ({\color{Mulberry}\ding{182}}). Requester first asks for content from its existing peers ({\color{TealBlue}\ding{183}}). If no peer has $C$, the requester uses DHT walks to find resolvers and fetch provider records ({\color{TealBlue}\ding{184}}). Requester connects to the provider and downloads $C$ via Bitswap (\ding{185}).}
  \label{fig:content-retrieveal}
\end{center}
\end{figure*}

IPFS is a content-centric system designed to shift file control and hosting from traditional cloud providers to P2P nodes, allowing other nodes to download content without specifying its locations. 

\paragraph{Content.}
When a file is added to IPFS, it is split into fixed-size chunks. 
These chunks become the leaf nodes of a Merkle-directed acyclic graph that collectively represents the self-verifying file structure. 
Each chunk is assigned a Content Identifier (CID), a self-certifying label generated by hashing the chunk’s content together with a small amount of metadata.
CIDs are location independent as they contain no information about where the content is stored, making them robust against node churn and network topology changes. 
They are also immutable (meaning, any modification of the content of a block results in a completely new CID), enabling versioning. 
Additionally, CIDs are not intended to be human-readable.
This design brings several advantages, including content deduplication (i.e., identical chunks across different files share the same CID), location-agnostic retrieval (i.e., a chunk can be fetched from any node that stores it, without depending on a specific host), and data integrity verification.

\paragraph{Publishing content.}

To add new content to the IPFS network, a node advertises itself as a provider of the CID representing that content via the DHT, see Figure~\ref{fig:content-retrieveal}.
First, the node creates a provider record, which binds its peer ID and the CID $C$ into a single record, indicating that it can serve the content on request.
Next, the node performs a DHT walk to locate the $k=20$ peers whose IDs have the smallest XOR distance to $C$, and pushes the provider record to these peers (step~{\color{Mulberry}\ding{182}}).
These peers, often referred to as resolvers~\cite{balduf2023cloud}, act as the lookup bridge between downloaders and providers.

For better robustness, IPFS also leverages ephemeral replication. When a node downloads content, it may temporarily act as a provider for several hours (duration depending on the local configuration) as the content remains cached.
If persistent hosting is desired, the downloader can publish its own provider record, effectively making itself a permanent provider for that content. This process of persisting and providing content is called pinning.
There are also dedicated pinning services, such as FileBase~\cite{filebase} and Pinata~\cite{pinata} that allow providers to store their content on other highly available P2P nodes for a fee, ensuring that these nodes also serve as alternative providers for the original content.

\paragraph{Retrieving content.}
Requesters always download the content from their connected peers via the Bitswap protocol~\cite{de2021accelerating}. 
In Bitswap, nodes exchange data based on two lists: a "want" list containing the CIDs that a node is seeking and a "have" list containing the CIDs that the node already possesses. 
Furthermore, nodes can also directly request CIDs from their peers' temporary "have" list (i.e., freshly cached content).
It is worth noting that Bitswap employs several strategies to promote cooperative behavior among peers.

Figure~\ref{fig:content-retrieveal} describes the process of finding the right nodes to download a given content. 
The process begins with the requester sending a query to all currently connected peers for the desired CID (see step {\color{TealBlue}~\ding{183}}).
This step can be completed in seconds, yet it is unreliable for unpopular or new CIDs, as they are unlikely to be populated widely in the network. 
If no connected peer responds with the content, the requester queries the DHT to look up to $k$ resolvers, i.e., nodes whose peer IDs are closest to $C$.
During this DHT walk, the requester also opportunistically asks encountered nodes for both the content and any relevant provider records.
The search stops once either 20 providers have been found or all $k$ resolvers have been queried, see step{\color{TealBlue}~\ding{184}}.
With the peer IDs of the content providers retrieved from the provider records, the requester then issues another DHT query to obtain their multiaddresses.
Finally, the requester establishes a connection to the chosen provider, and the content is transferred through Bitswap as in normal P2P exchange, see step~\ding{185}.

\subsection{Network-level Attacks}
\label{subsec:routing-attacks}

The Internet consists of over 77K individual networks, known as Autonomous Systems (ASes), which rely on the Border Gateway Protocol (BGP)~\cite{rekhter2006rfc} to exchange information on how to reach more than 1M IP prefixes~\cite{cidr}. 
Each AS originates one or more IP prefixes propagated to all other ASes.

\paragraph{BGP.}
BGP is a path vector routing protocol that determines how packets are forwarded to specific IP prefixes.
For a given prefix, the originating AS advertises it to neighboring ASes, which then propagate it hop-by-hop, each time prepending their own AS numbers, until the announcement reaches all ASes.
Each AS independently selects its best route toward the prefix and sets the next hop, with both route propagation and selection depending on the business relationships it has with neighbors.
More specifically, the two main relationships are customer-provider and peer-to-peer~\cite{as-relationship}.
In a customer–provider relationship, the customer pays the provider for full Internet connectivity.
In turn, the provider exports all its best routes to the customer and announces the customer’s prefixes to all its neighbors.
In a peer–to–peer relationship, two ASes connect to exchange traffic solely between their customer ASes and their own prefixes.
For route selection, an AS generally prefers customer-learned routes over peer-learned ones, and peer-learned over provider-learned ones.
Suppose that multiple routes to the same prefix are equally preferred (e.g., both are from customers). In that case, the AS chooses the one with the shortest AS path, breaking any remaining ties using arbitrary criteria (e.g., smaller AS number)~\cite{gao2001stable}.

\paragraph{Passive routing attacks.}
When packets travel between two computers, they often cross multiple ASes. 
Any AS along this routing path can observe network-level metadata, such as headers and packet sizes, and if the traffic is unencrypted, the content of the messages. 
Beyond packet inspection, a malicious AS can also delay or drop packets, turning this capability into a denial-of-service attack.
If the endpoints are IPFS nodes, such attacks can be used for content censorship. 
Importantly, a malicious AS does not need to manipulate the routing to intercept traffic from targeted victims, as it already lies on the routing path. 
We refer to this as a passive routing attack.
The effectiveness of such attacks depends heavily on the topological position of the malicious AS on the Internet, which determines the fraction of connections it can intercept naturally. 
Note that, due to the asymmetric nature of BGP routing, the attacker only needs to be on one side of the communication path, as the forward and reverse routes may traverse different Autonomous Systems (ASes) ~\cite{sun2015raptor}.

\paragraph{Active routing attacks.}

In BGP, route announcements are not authenticated by the ASes that propagate them.
This lack of authentication enables a malicious AS to inject bogus route announcements, commonly referred to as BGP hijacks, for one or more IP prefixes, causing other ASes to route traffic to an unintended destination \cite{sermpezis2018artemis}.
We consider such an event to constitute an active routing attack.
BGP hijacks can be broadly categorized according to the specificity of the bogus announcement relative to the legitimate one.
Suppose that the attacker announces a more specific prefix than the legitimate advertisement. In that case, the hijack will attract all traffic destined for that prefix, regardless of the attacker’s position in the Internet topology.
This is due to BGP's preference for the most specific matching route.
However, this technique is constrained by standard operational practices as network operators often filter advertisements for prefixes longer than /24 \cite{karlin2006pretty}.
If the attacker announces an equally specific prefix, the bogus route competes directly with the legitimate one.
In this case, the fraction of traffic diverted depends on the relative topological positions of the attacker and the victim.
For example, when the attacker-to-victim and origin-to-victim paths are similar in length, the adversary typically captures about 50\% of the traffic \cite{goldberg2010secure}.

Hijacks may also be distinguished by their operational complexity~\cite{sermpezis2018artemis}.
In an origin hijack, the attacker announces a prefix belonging to another AS without altering the origin AS field in the BGP announcement.
In contrast, a forged-origin hijack involves announcing the victim’s prefix while falsely listing the victim’s AS as the origin.
It is worth noting that the Resource Public Key Infrastructure (RPKI) \cite{rpki}, which cryptographically binds IP prefixes to their legitimate origin ASes, can effectively mitigate origin hijacks when widely deployed, but does not offer protection against forged origin hijacks.

\section{Censorship Attacks against CIDs}
\label{sec:attack}

In this section, we present the network-level censorship attacks against content retrieval in IPFS.
Essentially, the attack prevents requesters from downloading content from other IPFS nodes. 
We first introduce the threat model of network adversaries considered in this paper (\S\ref{subsec:threat-model}). 
We then describe the two main steps of the attack against given CIDs, namely reconnaissance (\S\ref{subsec:reconnaissance}) and censorship (\S\ref{subsec:censorship}).

\subsection{Threat Model}
\label{subsec:threat-model}

\paragraph{Attack capabilities.}
We consider a network adversary that aims to undermine the availability of content in the IPFS network. 
Specifically, we assume that the adversary controls a single Autonomous System (AS), referred to as the malicious AS. 
This AS can inspect and manipulate traffic traversing its infrastructure, e.g., by tampering with, dropping, or delaying packets. 
Beyond passive capabilities, the adversary can also mount active routing attacks, such as IP prefix hijacking through BGP manipulation (cf. Section~\ref{subsec:routing-attacks}).
Such capabilities are consistent with previous work targeting cryptocurrencies~\cite{apostolaki2017hijacking, saad2023three, tran2024routing}, Tor~\cite{sun2015raptor}, and PKI~\cite{birge2018bamboozling}, and are known to be within reach of ISPs, transit networks, or nation-state adversaries~\cite{nakibly2016website, wu2025wall}. 
A determined attacker may even establish a new AS and purchase transit from upstream providers for a few thousand dollars~\cite{setup-as}.
We also assume that the adversary operates an IPFS node to observe inbound peer connections. 
This node can be used to query the network for peers storing a specific CID. Running such a node without caching content is trivial, and hosting costs can be as low as a few dollars per month~\cite{ipfs-node-cost}.

\paragraph{Attack goals.}
The primary objective of the adversary is to censor content (i.e. CIDs) from the IPFS network. 
At scale, the adversary can attempt to disrupt the availability of a large volume of content simultaneously, effectively rendering entire services or platforms that rely on this content inaccessible. 
Alternatively, the adversary can target specific content accessed by particular user groups, often motivated by geopolitical interests (e.g., Spanish Internet censorship~\cite{spain-block-ipfs}), the need to block malicious or abusive material~\cite{unit42-ipfs-malware}, or efforts to enforce copyright protections~\cite{clouflare-block-ipfs}.

\paragraph{Assumptions.}

We assume a single adversary AS at any given time, as control over multiple ASes would only strengthen the adversary’s capabilities.
Thus, we leave a more comprehensive analysis of multiple adversaries for future work.
IPFS client nodes are generally hard to detect, as they are not added to the routing table of peers participating in the DHT. Therefore, the only way to observe IPFS client nodes is by passively waiting for incoming connections of IPFS client nodes to an IPFS server node. Being aware of this limitation, we explicitly do not take IPFS client nodes into our analysis and only consider the IPFS server node topology for our analysis.
For the analysis, we assume that the drift within the IPFS topology is small enough to be negligible, allowing for a static analysis over the entire measurement period. This is a standard assumption in the analysis of peer-to-peer networks, helping to reduce the complexity of the analysis.
We furthermore assume simplified BGP routing rules that always prefer shorter AS paths and maintain customer-provider and peer-to-peer relationships. In contrast, advanced BGP rules like local preference and BGP communities are not considered due to simplicity and a lack of transparency regarding the individual implementations of BGP routing rules at each AS. This assumption has been shown to accurately approximate the real routing of the internet for analysis of routing-based attacks ~\cite{tran2024routing}.


\subsection{Reconnaissance for Given CIDs}
\label{subsec:reconnaissance}

In the reconnaissance phase, the adversary seeks to identify all potential network-level surfaces that can be targeted for a given set of CIDs.
The output of this phase is a list of all relevant IPv4 and IPv6 prefixes of the network, which will serve as input for the next step.

\paragraph{Finding attack surfaces.}
Since a requester may retrieve content from existing peers or from the original providers, successful censorship requires disrupting the connections between the requester and these sources.
Another indirect approach to censoring content is to drop traffic between the requester and the resolvers responsible for provider discovery.
Consequently, the adversary must identify IPFS nodes that have cached the target content, the content providers, and the associated resolvers.

We show the reconnaissance steps in Algorithm~\ref{algo:reconnaissance}.
Specifically, to identify the IPFS nodes that have cached the target content, the adversary utilizes its own IPFS node and lifts any restrictions on its peering connections (Lines 6--8). 
In this way, the malicious IPFS node becomes a supernode with an unlimited number of peer connections. To increase the connectivity of this supernode, it could also actively connect to other IPFS nodes found during a DHT crawl. For the sake of comparability and simplicity, the attacker model does not include active connections to other IPFS nodes in the network for our analysis.
The adversary then queries all its peers for the target content by sending Bitswap want messages including the targeted CIDs. Then it resolves the IP addresses of those that have the content and maps these addresses with their corresponding prefixes and hosting ASes.  

Similarly, to learn IPFS nodes that have advertised themselves as provider resolvers, the adversary looks up the DHT for the 30 peers with the shortest XOR distance to each CID (line 4). Twenty peers are by default used as provider resolvers, however, newly joined peers could have entered among the 20 nearest peers since the last provider announcement. Therefore, we suggest taking 10 extra peers as a safety margin for the attacker. Furthermore, IPNIs configured by default are added to the provider resolver list. For this work, we considered the \emph{cid.contact} domain as IPNI used, as it is configured by default in Kubo v021.0.

By performing a standard CID lookup, the attacker queries all provider peer IDs and multiaddresses, including IPv4 and IPv6 addresses.
Lastly, the adversary also learns the prefixes and hosting ASes of these IP addresses (Line 9).

\begin{algorithm}[t!]
\caption{Reconnaissance for given CIDs.}
\begin{algorithmic}[1]
\STATE \textbf{Input:} List of CIDs
\STATE \textbf{Output:} ASes and  prefixes of Resolver peers, Bitswap providers, and Registered content providers
\WHILE{true}
    \STATE Extract IPNIs and provider resolvers by looking up 30 peers with the shortest XOR-distance to each CID
    \STATE Use IPFS nodes with unlimited peer connections:
    \STATE \quad Request all peers for targeted CIDs via Bitswap
    \STATE \quad Discover connected Bitswap Providers
    \STATE Request CID providers via a DHT search
    \STATE Lookup AS number and prefixes and store results:
\STATE \quad \texttt{resolver\_prefixes} $\leftarrow$ resolver peers
\STATE \quad \texttt{bitswap\_prefixes} $\leftarrow$ bitswap providers
\STATE \quad \texttt{cid\_provider\_prefixes} $\leftarrow$ content providers
\ENDWHILE
\end{algorithmic}
\label{algo:reconnaissance}
\end{algorithm}

\paragraph{Measuring attack surface on live IPFS network.}
\label{par:measure-attack-surface}

In previous work like \cite{balduf2022monitoring}, the collection of CIDs was performed based on Bitswap logs. 
This approach has proven to be well-suited to explore CIDs that are currently in the scope of requesters. 
Thus, the same approach was used to collect the CIDs for this work. 
It was also important to note that, in addition to the CIDs themselves, the registered providers, as well as the peers that cache the content, are retrieved. 
Therefore, it is crucial to minimize the delay between the collection of CIDs and the lookup of their content providers and caching Bitswap providers. 
Hence, a minimal delay is achieved by utilizing a rolling window, implemented as a FIFO queue, which stores recently requested CIDs along with their corresponding peer IDs. 
This rolling window enables the selection of CIDs that are frequently requested by multiple independent requesters and are therefore assumed to be of high relevance at a given time.
%
To retrieve CIDs from Bitswap logs, as well as to find its provider and resolver IPFS nodes, we use a setup extending the Bitswap monitoring infrastructure of ~\cite{balduf2022monitoring}. 
The base infrastructure consists of two Bitswap monitoring IPFS nodes, a RabbitMQ server, and a Prometheus server. 
The Bitswap monitors are standard Kubo v0.21.0 IPFS nodes configured to accept an unlimited number of connections and utilize a custom Bitswap plugin that logs all incoming Bitswap messages to a RabbitMQ queue. 
Furthermore, the plugin allows one to send Bitswap wantlists to the node's peers actively.
This enables us to query our peers for specific CIDs actively.
Our extension to this base infrastructure includes five additional Bitswap monitoring nodes, where one of the five also allows for an unlimited number of connections. 
In comparison, the other four nodes kept the default connection limits. 
The whole infrastructure, together with all clients, runs on a CX52 virtual machine at Hetzner GmbH running on an Intel Xeon physical CPU with 16 shared vCPUs and 32 GB of RAM.

For the collection of CIDs, the incoming wantlists from all Bitswap monitoring nodes are read from the RabbitMQ queue, and for every CID in the wantlist, the CID is stored together with a set of all requesting peers into a FIFO queue of size 10'000. 
For CIDs already found in the FIFO queue, the peer ID is attached to the set of peers; otherwise, the CID is appended with a fresh peer ID set. 
To select relevant CIDs, each CID in the FIFO queue is scored with a weighted sum of its peers, where the weight of each peer is inversely proportional to how often the same peer occurred in other CIDs. 
This score aims to find CIDs that are relevant to many peers while focusing on peers that are less noisy and more specific about the choice of CIDs they are interested in. 
During selection, the CID with the highest score is evaluated by looking up its providers in the DHT using the IPFS RPC API of the second Bitswap monitor. 
If no providers could be found within a timeout period of 2 seconds, the CID was not further considered otherwise, all provider nodes found are stored with their multiaddresses. 
Additionally, to capture IPFS nodes that cache content and could serve it directly via Bitswap, the first of the five added Bitswap nodes with unlimited connections is queried for Bitswap providers using the Bitswap plugin. 
This node is assumed to be attacker-owned, and the found IPFS peers are stored together with their multiaddresses. 
Since the attacker does not have a full view of the network, it is also essential to collect caching IPFS nodes from the perspective of independent IPFS nodes, which are considered attack targets.
This is achieved by querying the same CID also from the four remaining Bitswap nodes with default connection limit configurations. 
The found caching peers are also stored, together with their multiaddresses. 
For all the provider peers found, all exposed IPv4 and IPv6 addresses were retrieved, and the IP prefixes, along with their AS numbers, were extracted using the pyasn library ~\cite{pyasn162} and a dataset downloaded via the pyasn library \texttt{pyasn\_util\_download.py --latestv46} command on June 24, 2025. In total, three datasets consisting of 1,000 CIDs were collected in July 2025, with a one-week gap between each collection. The collection of a single dataset takes approximately one day.

\paragraph{Findings.}

\begin{figure}[t!]
\begin{center}
  \includegraphics[width=0.48\textwidth]{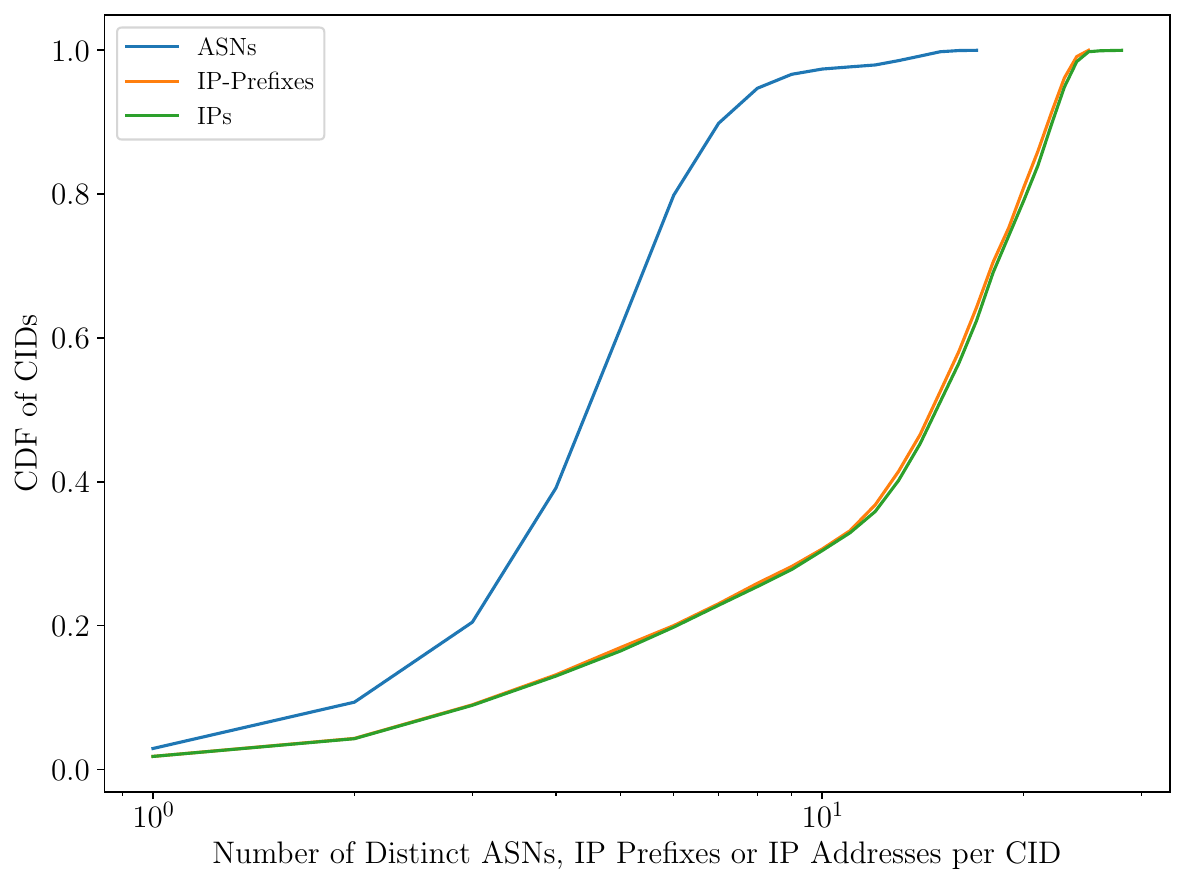}
  \caption{Shows the cumulative distribution function for the number of distinct ASNs, prefixes, and IPv4 and IPv6 addresses for the combination of provider and caching Bitswap providers in the three CID datasets, each consisting of 1000 CIDs. Most of the collected CIDs are located in fewer than 5 ASes and fewer than 20 IP prefixes, which indicates a high concentration on a few provider nodes.}
  \label{fig:popular_cid_provider_bitswap_cdf}
\end{center}
\end{figure}

\begin{figure}[t!]
\begin{center}
  \includegraphics[width=0.48\textwidth]{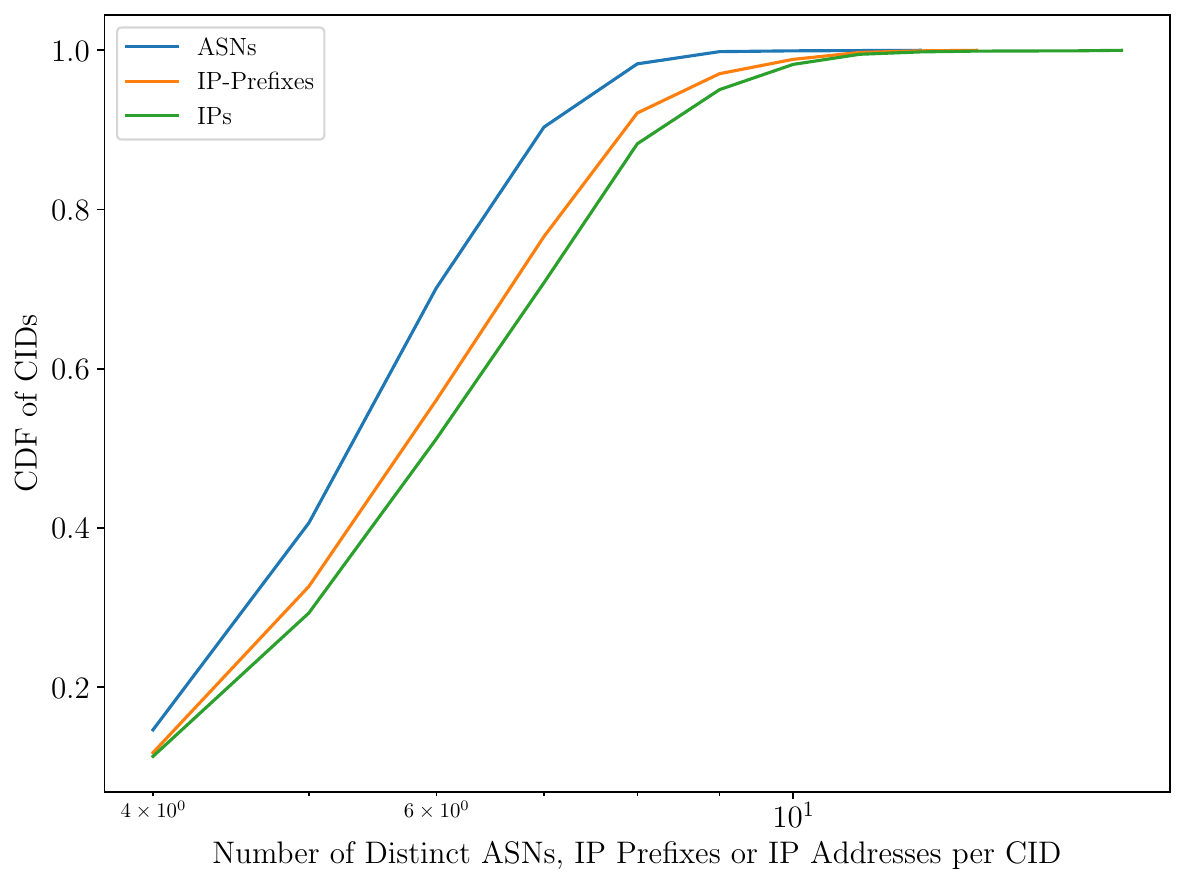}
  \caption{Shows the cumulative distribution function for the number of distinct ASNs, prefixes, and IPv4 and IPv6 addresses for the combination of provider and caching Bitswap providers in the three CID datasets, each consisting of 1000 CIDs. Interestingly, even though there are 20 resolvers, many of these resolvers are hosted on the same IP address or prefix, leading to a higher centralization than the replication count of 20 would suggest.}
  \label{fig:popular_cid_provider_record_bitswap_cdf}
\end{center}
\end{figure}

The three datasets were fused and analyzed for their centralization in terms of distinct IP addresses, IP prefixes, and ASes for both the first attack vector, consisting of the content provider together with the caching Bitswap provider nodes, and the second attack vector, composed of content resolvers together with the caching Bitswap providers. 
The results of the analysis are shown in \ref{fig:popular_cid_provider_bitswap_cdf} \ref{fig:popular_cid_provider_record_bitswap_cdf}. 
Both the content providers and the resolvers show a centralization in a few ASNs and IP prefixes. 
At the same time, there is also a small fraction of CIDs that are well distributed over a larger number of ASes (10+) and IP prefixes (20+). 
Another interesting observation is that even if for the resolvers, a replication over approximately 20 would be expected due to the replication over 20 peers, the total number of resolvers' IP addresses is often way below this number, which can be attributed to the large number of IPFS server nodes being hosted behind a NATed IP address.

\subsection{Routing-based Censorship}
\label{subsec:censorship}

BGP routing-based attacks on providers can be conducted either by passive interception or active BGP hijacking. 
Both types of attacks have their unique advantages. 
Although passive interception does allow for intercepting traffic without the need for BGP manipulation, which makes it much more stealthy, it is only possible to attack connections that naturally route through the attacker's AS. 
BGP hijacking attacks can, on the other hand, be very effective, especially against IP prefixes that are not adequately protected.
However, the need to announce bogus IP prefixes by the attacker makes them relatively easy to detect and mitigate. 
For this reason, we evaluate both types of attack, as well as their combination, for our routing-based censorship.

The first type of BGP routing-based attacks includes passive BGP interceptions, where an AS can attack a connection by controlling an AS that is on the AS route picked by BGP to send traffic from the requester to the provider or the AS route from the provider to the requester. 
The forward and backward paths in general do not need to be the same, as they are dynamically determined based on the topology by the BGP protocol. 
In general, it is hard to protect against this type of attack, as BGP does not allow active route selection. 
 So, the only way to counteract passive BGP traffic interception is to reconfigure BGP rules so that the intercepting AS is circumvented.
The calculation of AS paths required to find passively intercepting ASes is a well-studied problem. 
For this work, we based the evaluation on the implementation of ~\cite{vonarx2023revelio}. 
It builds an AS network topology based on AS relationships ~\cite{caida_as_relationships} and IXP datasets ~\cite{caida_ixps_dataset} published by CAIDA. 
Based on the topology, a routing tree is derived based on the customer, peer, and provider relationships. 
The shortest AS-paths further filter competing paths, and the ties are broken by choosing smaller AS numbers. 
The algorithm for calculating the routing tree is taken from ~\cite{goldberg2010securebgp}. 
Based on the routing trees, the paths between any two ASes can be calculated by traversing the routing tree from the root, which represents the target, to the requester.

In comparison to the passive BGP interception attacks, active BGP Hijacking involves announcing bogus BGP announcements from a malicious AS to claim a route towards the provider IP prefix and thereby attracting traffic from the requester towards the provider. 
While in principle it is also possible to attack the traffic towards the requester, this approach is often less scalable for a large number of requesters, as BGP hijacking, especially against many targets, can be detected much more easily and comes with a high load on the attacker's bandwidth. 
For our work, therefore, we only consider the hijacking of provider IP prefixes.
For the analysis of BGP hijacking attacks, the IPv4 or IPv6 prefix was first evaluated for its RPKI status. The RPKI status was retrieved by a client calling the irrexplorer endpoint ~\cite{irrexplorer2025} to sort each BGP prefix into one of the following four categories:
\begin{enumerate}
  \item Not RPKI-protected prefix length smaller than /24 (IPv4) /48 (IPv6)
  \item Not RPKI-protected prefix length is /24 (IPv4) /48 (IPv6)
  \item RPKI-protected prefix length is shorter than the max-prefix length specified in the RPKI entry.
  \item The RPKI-protected prefix length is equal to the maximum prefix length specified in the RPKI entry.
\end{enumerate}

For both categories 1 and 2, the attacker can announce the origin without being filtered out by RPKI checks. 
In the first case, an attack can hijack all connections by announcing a /24 prefix containing the target IPv4 address. 
In the second case, it will compete based on the AS length, where the shorter AS length wins. 
In the case of active RPKI protection, the only way to hijack traffic is to use a valid announcement and attach itself to the path. 
In case 3, by announcing the maximum prefix length, the traffic can still be hijacked for all ASes.
In case 4, the only chance for the attacker to win is to be at least one hop closer to the target than the legitimate provider. 
Based on these rules, Cases 1 and 3 can be counted as hijacked directly, while for Cases 2 and 4, the intercepting ASes were searched in a breadth-first search around the requesting node. \\

\begin{figure}[t!]
\begin{center}
  \includegraphics[width=0.45\textwidth]{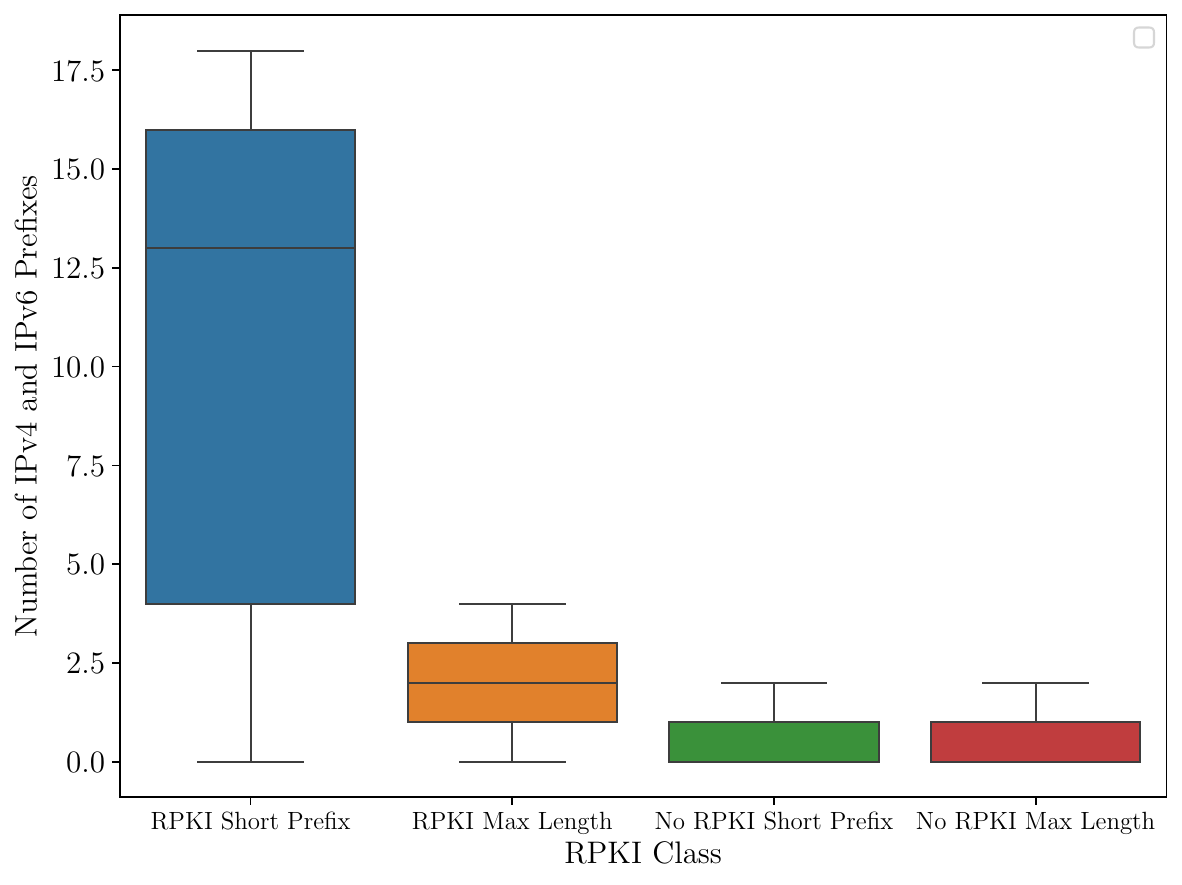}
  \caption{Shows the number of unique IP prefixes of all content providers together with caching Bitswap providers for the combination of the three datasets, consisting of 1000 CIDs each. Interestingly, the vast number of IP prefixes is in the RPKI short prefix category and therefore unprotected against BGP hijacking. Only a small number of prefixes fulfill max prefix lengths and are therefore better hardened against BGP hijacking.}
  \label{fig:popular_cid_provider_bitswap_rpki}
\end{center}
\end{figure}

\begin{figure}[t!]
\begin{center}
  \includegraphics[width=0.45\textwidth]{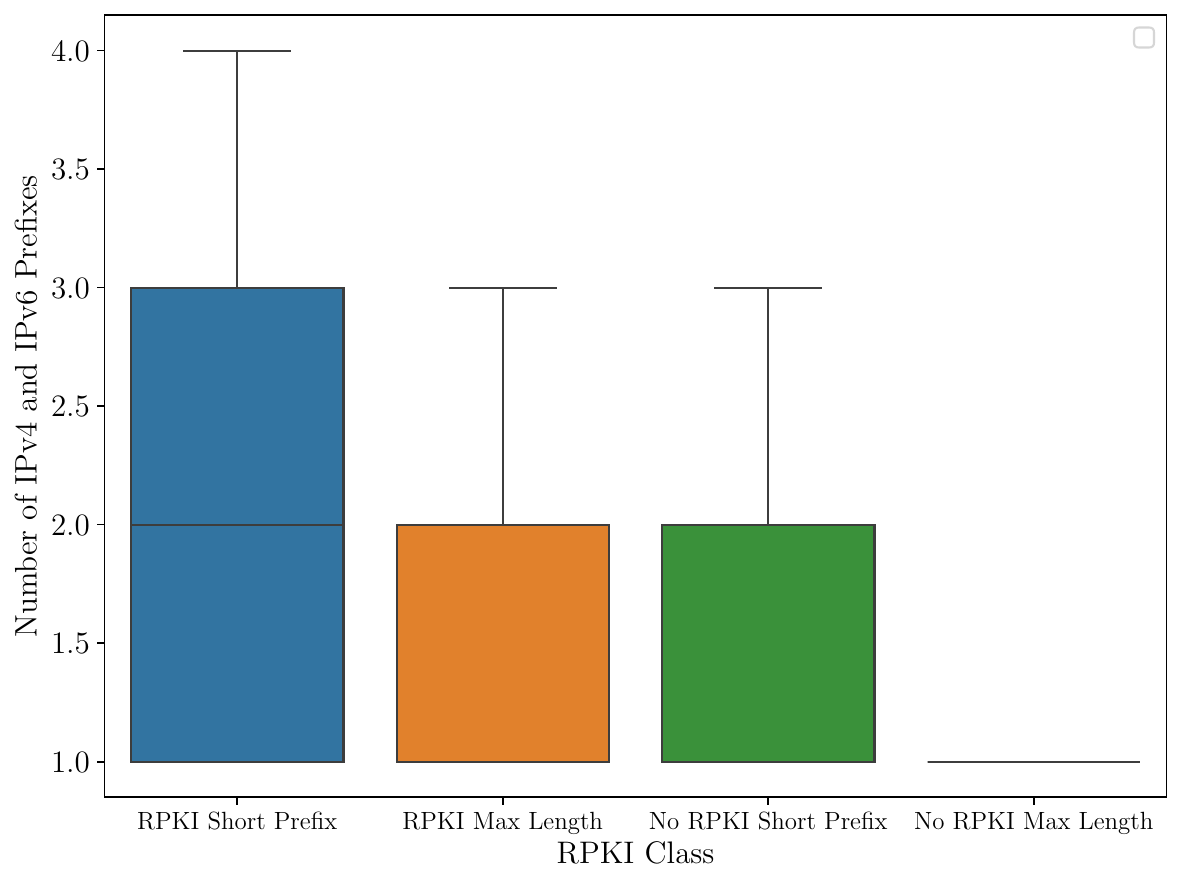}
  \caption{Shows the number of unique IP prefixes of all provider record resolvers together with caching Bitswap providers for the combination of the three datasets, consisting of 1000 CIDs each. Interestingly, there are always at least two prefixes in a well-protected category \emph{RPKI Max Length} or \emph{No RPKI Max Length}, making it hard for an attacker to block all prefixes with BGP hijacking.}
  \label{fig:popular_cid_provider_record_bitswap_rpki}
\end{center}
\end{figure}

An analysis of the combination of the three CID datasets, consisting of a total of 3000 CIDs for their distribution in RPKI protection for both attack vectors, the provider nodes together with the caching Bitswap providers, and the resolver nodes together with the caching Bitswap providers, is shown in Figure \ref{fig:popular_cid_provider_bitswap_rpki} and Figure \ref{fig:popular_cid_provider_record_bitswap_rpki}. 
Interestingly, for both attack vectors, the vast majority of IP prefixes fall into category 1 or 3, rendering them vulnerable to BGP hijacking. 
A further important observation is that the provider attack vector generally has slightly less well-protected IP prefixes in categories 2 or 4, and that the provider record attack vector always has at least one prefix in both categories 2 and 4. 
This observation suggests that the provider attack vector is better suited for BGP hijacking.

\section{Evaluation}

\paragraph{Methodology.}
To determine the topology of the IPFS nodes in the network that could potentially act as requesters, the IPFS crawler by~\cite{balduf2023cloud} was used. 
It scans the IPFS network by generating random peer addresses in each DHT bucket of a peer and finds the closest peer to it. 
Thereby, the routing table of the peer can be queried exhaustively, and unseen peers can be recursively scanned in the same way.
Utilizing this crawling method, a complete graph of the connections in the IPFS network between IPFS server nodes can be determined at any given time. 
The full search only takes around 5 minutes, which keeps the topology drift during the measurement small. 
To build the requester dataset, two IPFS crawls were executed on July 12 and 13, 2025. 
The snapshots were concatenated and deduplicated by IPv4 and IPv6 addresses. 
Two snapshots were taken to ensure that nodes that might not have been reachable in one of the scans are still included in the dataset. 
The deduplication by IP address helps counteract an overestimation of the importance of services that run several IPFS nodes behind a single IP address. 
A limitation of this approach is that IPFS client nodes cannot be observed using the IPFS crawler, as they are not part of the routing tables of IPFS server nodes. 
For each requester, the same approach as for providers was used to retrieve the AS and IPv4 and IPv6 prefixes associated with the requester's IPv4 and IPv6 addresses \ref{par:measure-attack-surface}. As IPFS API gateways run regular IPFS server nodes as well, they are part of the requester set. 
Similar to other downstream services of other IPFS nodes with potentially many more consumers, gateway IPFS requester nodes are not weighted by actual consumers.
Therefore, the results based on this requester dataset are limited to the availability of content towards IPFS server nodes. 

For the attacker ASes, only the 100 highest-ranked ASes according to~\cite{caida_asrank} were considered for the analysis; for the complete list, consult Appendix~\ref{appendix:top-rank-ases}. 
Generally, a higher rank indicates better connectivity of the AS to other high-rank ASes, placing it in a generally favorable position for both passive interception and BGP hijacking. 
This limitation of considered attacker ASes is put in place to limit the computational complexity of the simulation and to also not overfit on obscure ASes, and instead keep it to ASes that are well within reach of a nation-state actor or ISP. 
For the case where one of the four victim-owned IPFS nodes observed a caching Bitswap provider that could not be found by the attacker IPFS node (see Section~\ref{par:measure-attack-surface}), we assume the attacker to have failed to block the CID, which results in no requester being blocked in our analysis, as the attacker would not be able to entirely block all caching Bitswap providers found by IPFS nodes in the network. 
However, this case was quite rare and occurred in only 107 out of the 3,000 CIDs in the three datasets.

\subsection{Attack Feasibility}

\begin{figure}[t!]
\begin{center}
  \includegraphics[width=0.45\textwidth]{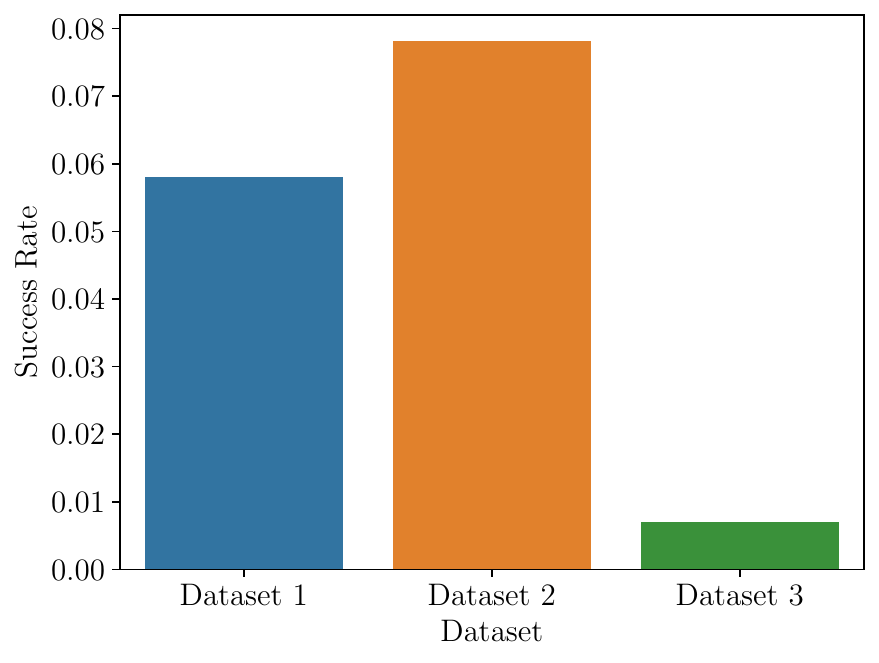}
  \caption{Shows fraction of CIDs that can be fully blocked for all requesters per dataset. The comparison shows that for all datasets, only a small fraction of CIDs can be fully blocked, with a significant variance between the three datasets.}
  \label{fig:all_providers_blocked}
\end{center}
\end{figure}

A primary research objective was to determine whether, in our three collected CID datasets, each consisting of 1000 CIDs, there are any CIDs that can be blocked from all requesters, and if so, for which fraction of CIDs this complete blockage is possible.
As shown in Figure \ref{fig:all_providers_blocked}, for all three datasets, a small fraction between $0.5\%-8\%$ of all CIDs can be blocked entirely from all requesters. 
Interestingly, we observe significant differences between the three datasets.
Both a small sample size and a dependence between the samples can explain the considerable difference between the measurements. 
A further analysis of the samples that could be blocked entirely revealed that all of them lacked a provider protected against BGP hijacking, either by having a max-prefix RPKI-protected route or a prefix length of 24 for IPv4 or 48 for IPv6. 
This makes the prefixes naturally vulnerable to BGP hijacking and does not restrict the attacker AS to be close to the requester. 
As shown in Figure \ref{fig:popular_cid_provider_bitswap_cdf}, especially RPKI-protected prefixes with short prefixes make up a large number of provider prefixes, making them vulnerable to BGP hijacking.

\subsection{Attack Effectiveness}

\begin{figure}[t!]
\begin{center}
  \includegraphics[width=0.45\textwidth]{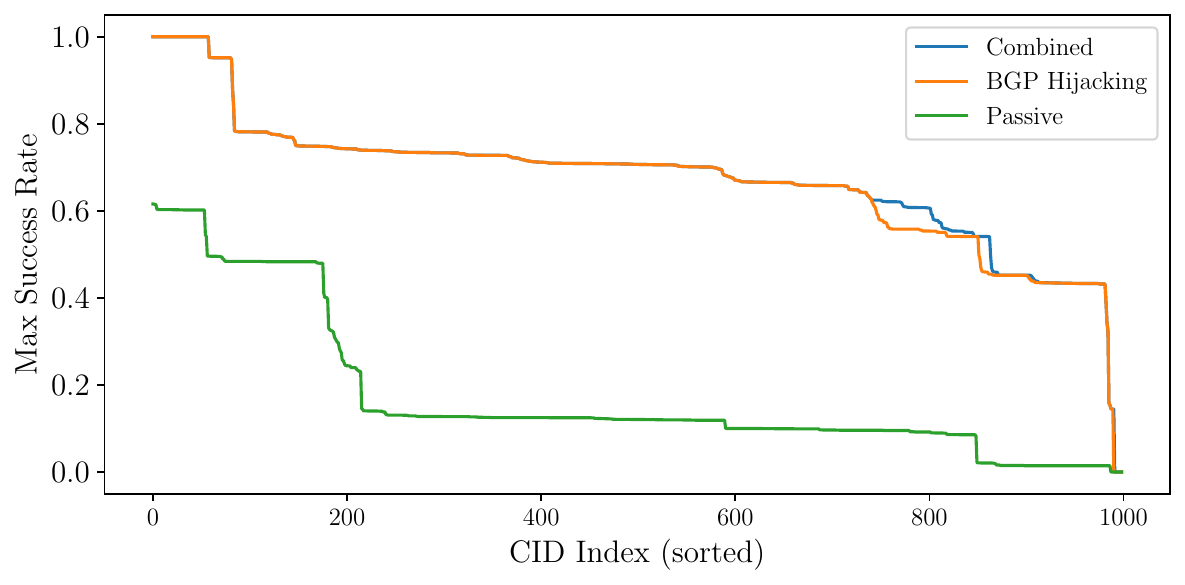}
  \caption{Shows the maximal rate of successfully blocked requesters by a single AS, denoted as the maximal success rate for dataset 1. The plot compares BGP hijacking, passive interception, and their combination. With passive interception, only a small fraction of requesters can be intercepted for most CIDs. At the same time, BGP hijacking generally proves to be very effective as it allows for blocking over 70\% of the requesters for most CIDs.}
  \label{fig:partition_providers_blocked_attack_types}
\end{center}
\end{figure}

Next to be evaluated are the CIDs that cannot be entirely blocked from all requesters, if they can at least be blocked from a large majority of requesters. 
The results shown in Figure~\ref{fig:partition_providers_blocked_attack_types} show the maximal fraction of requesters that a single AS can block. 
For most CIDs, passive interception can only block less than 20\% of all requesters, while with BGP hijacking, a 70\% blockage rate can be achieved for most CIDs. 
For some CIDs, the combination of BGP hijacking and passive interception results in a slightly higher blockage rate.

\begin{figure}[t!]
\begin{center}
  \includegraphics[width=0.45\textwidth]{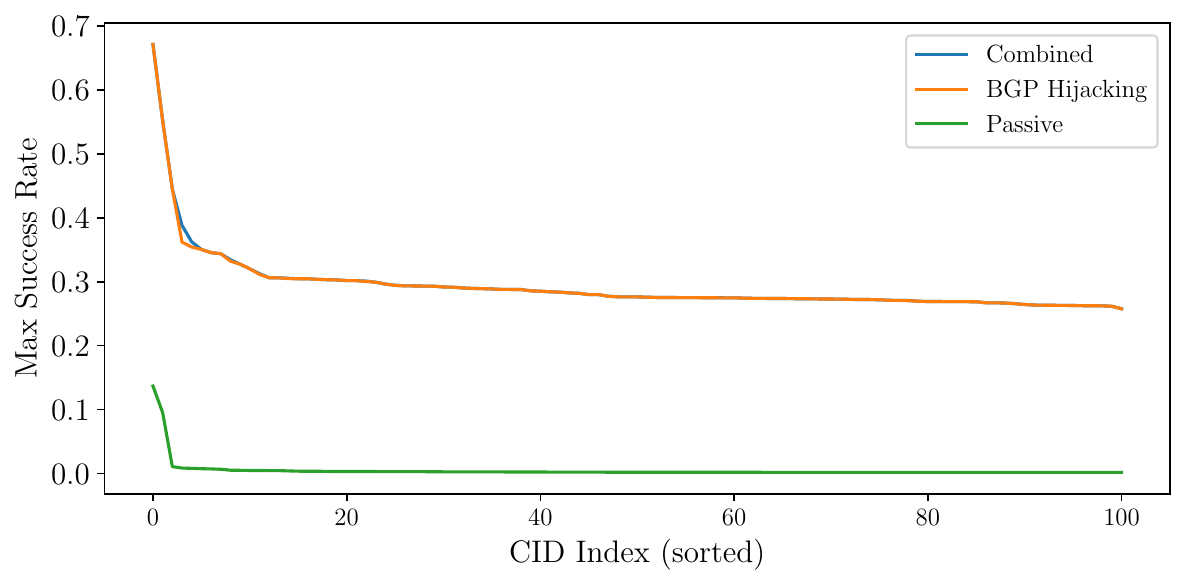}
  \caption{Shows fraction of CIDxRequester pairs that a single AS can block for the 100 highest ranked ASes. The plot shows that there exist a few very powerful ASes AS174, AS1299, and AS1031 that can block 67, 55, and 44\% of all CIDxRequester pairs, while most other ASes can only block around 30\% of all CIDxRequester pairs. This shows that a small set of highly important ASes is, on average, able to launch much more effective routing attacks against IPFS content.}
  \label{fig:as_overview_blockage_rate}
\end{center}
\end{figure}

In Figure \ref{fig:as_overview_blockage_rate}, the blockage rate for all CIDs in dataset 1 is displayed for the 100 ASes with the highest ranking. 
Interestingly, 3 ASes are capable of blocking a higher fraction of requesters on average. These ASes are AS174, AS1299, and AS1031 with 67, 55, and 44\% average blocking rate with BGP hijacking. 
At the same time, most of the other top-ranked ASes only reach a blockage rate of around 30\%. 
As a result, we conclude that the effectiveness of BGP hijacking to censor content is very dependent on the adversary ASes, where strongly connected ones like AS174 clearly have an advantage.

\begin{figure}[t!]
\begin{center}
  \includegraphics[width=0.45\textwidth]{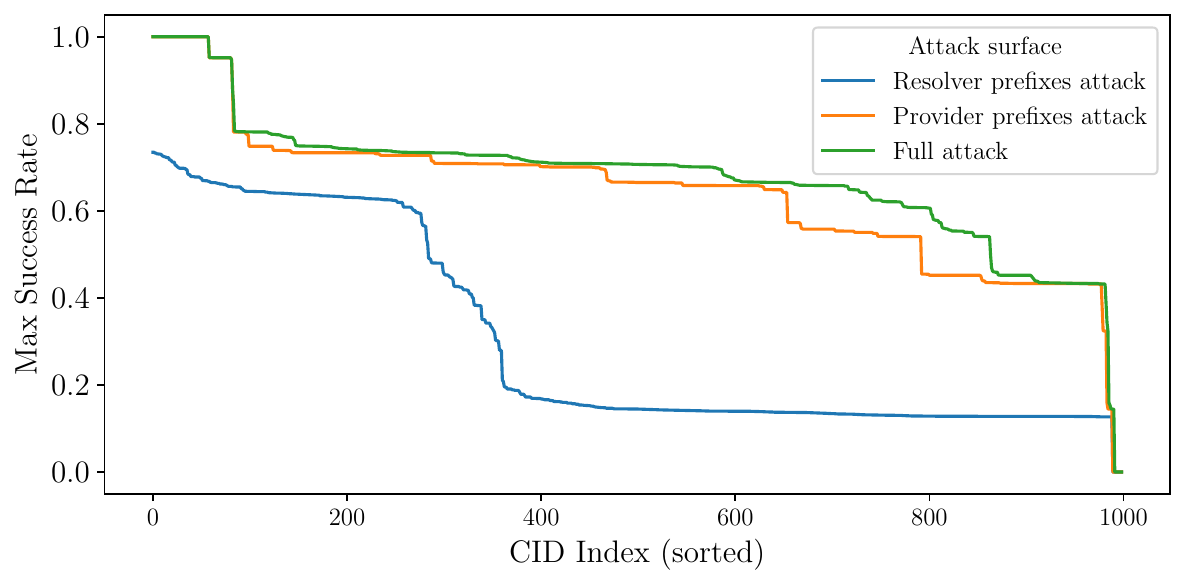}
  \caption{The maximal rate of successfully blocked requesters by a single AS in dataset 1. 
  Attacks against providers are usually far more effective, resulting in rates of over 70\% of requesters being blocked for most CIDs. 
  }
  \label{fig:attack_vector_comparison}
\end{center}
\end{figure}

In Figure \ref{fig:attack_vector_comparison}, the different attack vectors of blocking the content providers together with Bitswap providers or of blocking the content provider resolvers together with the Bitswap providers are compared with the full attack that combines the two attack surfaces. 
In general, the attack on the resolvers is weaker, while adding the resolvers increases the total fraction of blocked requesters. 
However, the difference in the fraction of blocked requesters between the full attack and the partial attack is often insignificant.

\subsection{Attack Cost}

BGP hijacking usually comes with a cost for an attacker in terms of easier detectability and more evident signs of malicious intent. 
An attacker, therefore, strives for the maximal blockage with the minimum cost in terms of IP prefixes that need to be hijacked.

\begin{figure}[h!]
\begin{center}
  \includegraphics[width=0.45\textwidth]{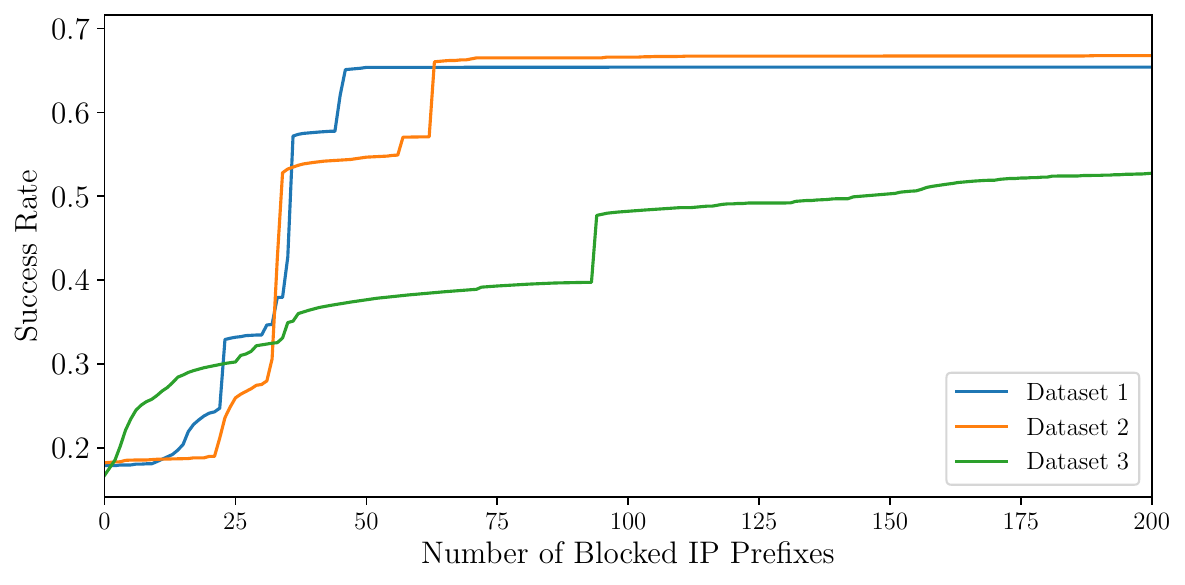}
  \caption{The fraction of blockable requesters is on average for the different datasets, when an attacker can use a certain number of hijacked IP prefixes. 
  It shows that even a small number of prefixes can already block the majority of hijackable requesters.}
  \label{fig:cost_comparison}
\end{center}
\end{figure}

To determine the fraction of requesters that can be blocked on average over all CIDs with a given budget of hijackable IP prefixes, a greedy algorithm was applied. 
The algorithm prefers prefixes with the most significant immediate gain. 
In the event of no immediate gain, prefixes are added that require the fewest additional prefixes until a subsequent gain is achieved.
Ties are broken by iterating over the CIDs in the order of the datasets, and the current best prefix is only overwritten if it is better than the previous best prefix found. 
Although in general this algorithm will not find the optimal set of prefixes, it gives an efficient calculation for a lower bound of what an attacker can achieve with the given number of hijackable prefixes. 
Figure \ref{fig:cost_comparison} shows the results of this evaluation for all three datasets. 
The results show that a small number of prefixes (50-100) out of a total of 2791 prefixes is enough for the attack to reach almost maximal attack efficiency.
This indicates the attack's feasibility in practice.



\section{Countermeasures}

\subsection{Vulnerability evaluation of existing IPFS clusters}
\label{subsec:countermeasures-vulnerability-evaluation-existing-ipfs-clusters}


IPFS clusters aim to allow private or public clusters of IPFS nodes to collaboratively pin a set of documents and reach a configurable replication count~\cite{ipfs-cluster}. 
Especially when IPFS nodes within the IPFS cluster are spread over different ASes and IP prefixes, IPFS clusters are a promising countermeasure against passive interception and BGP hijacking attacks. 
In a first step, the three open IPFS cluster projects \emph{Project Gutenberg (Spanish)}, \emph{IPFS Websites}, and \emph{Wikipedia} posted on the collab.ipfscluster.io website \cite{ipfs-cluster} were analyzed for their ASes and IP prefixes, see Appendix \ref{appendix:countermeasures-collab-as-prefixes}. 
All of them had providers in AS54825 that had a max-length RPKI-protected prefix, which makes the CIDs extremely difficult to attack with a BGP hijacking attack. 
A later simulation of our attack against 1000 CIDs collected from each project was unsuccessful in blocking any requesters from accessing the CIDs. 
Although in this case it cannot be attributed to the distribution around many ASes and IP prefixes, it shows that a very effective countermeasure against our attack is to use at least one provider node that uses an IP address in an IPv4 and IPv6 prefix range that is RPKI-protected and announces the max-length prefix. 
By utilizing several of these well-protected prefixes, the provider can further harden against traffic interception, as passive interception will also become more challenging when more than one provider in different parts of the internet topology must be blocked.

\subsection{Global random pinning collaboration}
IPFS clusters, combined with well-protected IP prefixes, provide a robust method for content providers to safeguard their content. 
However, this shifts the responsibility for securing content in IPFS to the providers, which does not align with the goal of IPFS to make all content in IPFS censorship-resistant. 
To address this issue, we therefore propose a concept similar to IPFS clusters but on a global level, with direct integration into the main IPFS protocol. 
The main idea is that all nodes can share content in the network, accompanied by a replication parameter that, by default, is set to a value that reasonably protects the content against routing-based attacks. 
When requesting CID pinning for a document in turns, one has to host the same amount of storage $\times$ the replication rate for other participants. 
There would of course be challenges in terms of ensuring that nodes that do not comply with storing content for others are punished, e.g., their content is dropped. 
However, the exact protocol for ensuring that all participating nodes comply with the protocol without abusing the system to engage in selfish pinning is left as an open question. 
Our contribution in analyzing such collaborative pinning in IPFS lies in testing the randomized pinning among all IPFS server nodes. 
Our assumption for the analysis is that all IPFS server nodes pin a certain CID with the same probability.

\begin{figure}[t!]
\begin{center}
  \includegraphics[width=0.45\textwidth]{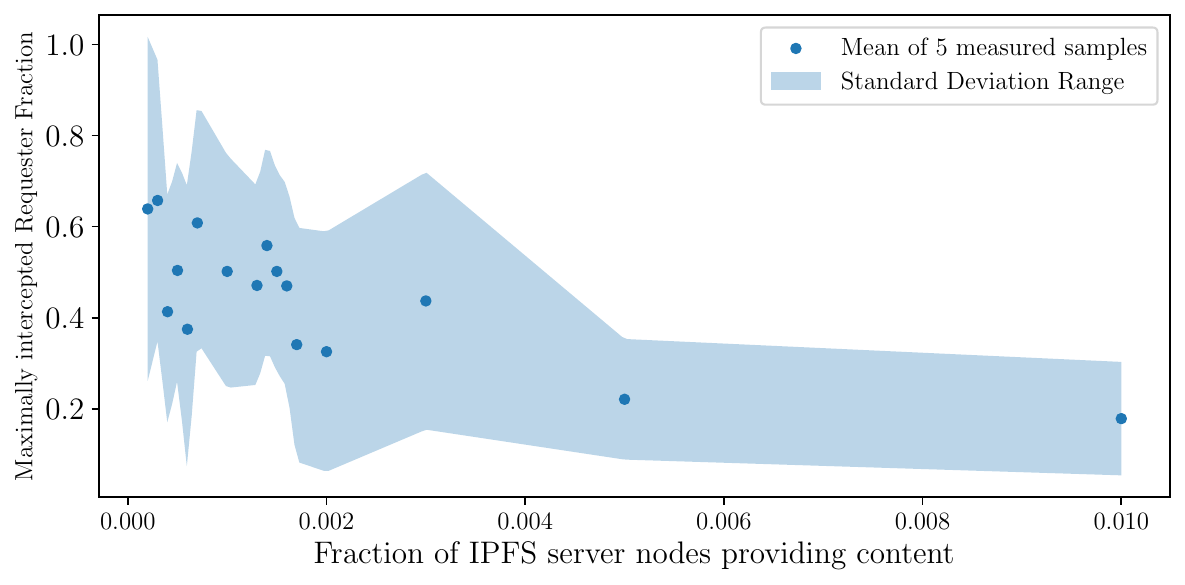}
  \caption{Shows for different fractions of all IPFS server nodes that participate in a global collaborative pinning of a CID, the fraction of requesters that can be maximally intercepted by one of the 100 highest-ranked ASes using a combination of BGP hijacking and passive interception. 100\% corresponds to 16107 nodes. After around 0.5\% of all IPFS server nodes, corresponding to around 80 nodes, the attack effectiveness converges to around 20\% }
  \label{fig:public_ipfs_cluster_random_joining}
\end{center}
\end{figure}

As Figure~\ref{fig:public_ipfs_cluster_random_joining} shows, a replication of content to 0.5\% of IPFS server nodes, which corresponds to 80 nodes, is enough to limit the interception rate to 20\%. 
As we see from the analysis in \ref{subsec:countermeasures-vulnerability-evaluation-existing-ipfs-clusters}, well-protected nodes that are RPKI-enabled and announce the prefix according to the RPKI max length can further boost the protection of content. 
Adding strong nodes to the system that replicate most content and provide it over a well-protected IP prefix could further improve the security posture of content in the network, possibly bringing the interception rate down to almost 0. 
This combination is presumably the most resilient, as it combines strong protected nodes with substantial decentralization of the collaborative pinning, making it more resistant to interception attacks and DDoS-type attacks.
\section{ Discussion }
\label{sec:discussion}


\paragraph{Requester selection.}
Generally, a limitation of our evaluation is that we take a simplified approach to collecting the requester set by utilizing a network crawl and deduplicating requesters based on their IPv4 and IPv6 addresses.
This approach fails to fully capture the complete requester set, as IPFS client nodes and multiple IPFS server nodes behind a NAT are not fully accounted for. 
Furthermore, since IPFS nodes typically act as intermediary providers of content, different IPFS nodes may, in turn, serve very different requester sets, each with its own size and importance to an attacker. 
So, one IPFS node may serve millions of end-users, while another IPFS node is not even used to serve content at all. 
This difference in importance is not captured in our evaluation. 
Possible ways to refine the review in the future would be to utilize Bitswap requests and other telemetry to investigate the importance of each IPFS node further and take this information into account during the evaluation.

\paragraph{CID selection.}
Another limitation to note is that the selected CIDs in each dataset, although based on requests from different peers, do not guarantee any real-world relevance. 
Furthermore, the impact of blocking each of these CIDs can vary significantly from one CID to another. 
A more purpose-based CID collection method might here give a more complete picture of how vulnerable CIDs are, depending on their use case.

\paragraph{Model approximations and real-world considerations.}
Although the modeling of BGP hijacking and passive interception utilizes well-tested principles that provide a good approximation of reality, it still misses specific topological details of BGP implementations. 
Furthermore, the idealized state of BGP we assume for our attack might change very quickly if an AS is assumed to perform malicious actions and other ASes start to mistrust it. 
Therefore, our analysis, even considering the cost implications, is rather an upper bound of what an attacker can achieve, at least if the attack is not globally legitimized, for example, by rule makers censoring certain content. 
Although global attacks on content are therefore rather unlikely, the use of this attack type to do local censorship, for example, restricted to one country, is, however, as shown in this work, very well possible if the provider does not take specific counteractions against it. 
In cases where the attacker does control ASes on all paths to all requester nodes, for example, a nation-state actor in its own country, IPFS in itself does not give any direct protection. 
The only chance would be to use proxies or secure tunnels into unrestricted areas, utilizing, for example, virtual private networks (VPN) or Onion Routing (eg, Tor) if these are possible.
\section{Related Work}


\subsection{Measurements of IPFS network}
\label{subsec:rel-measurement}
As IPFS like other decentralized networks is not fully controlled by a single entity but instead grows dynamically based on the interaction of many parties and further interact with technological and economical trends like, for example, cloud adoption, an important question is if the real-world IPFS network lives up to the performance, scalability and security characteristics, the maintainers of IPFS are aiming to achieve. 
To answer this question, numerous prior publications have focused on methods for measuring different aspects of IPFS in the real network.
In \cite{crawling}, a crawling technique was introduced that several works, including ours, built on \cite{balduf2022monitoring, henningsen2020mapping}. 
It allows crawling the DHT and thereby mapping out IPFS server nodes currently active in the IPFS network. 
\cite{henningsen2020mapping} could thereby, through repeated crawls, show that the IPFS network on average had 44,000 nodes, of which, however, only a fraction was reachable. Furthermore, at the time of measurement, most nodes in the IPFS network were found to be operated by private individuals. 
Another method of acquiring topology information is to use standard IPFS server nodes with default or unlimited connections or Hydra nodes to monitor incoming peer connections \cite{henningsen2020mapping, daniel2022passively}. 
Although this method allows only for a local view, it comes with the advantage that it makes client nodes observable, which, for IPFS crawlers, are not conceivable due to their absence in the DHT. 
Since IPFS nodes respond only to requests containing the correct CID, exploring a CID dataset that resembles the CIDs in the network is not trivial. 
To overcome this challenge, various works, including ours, have utilized Bitswap messages to discover CIDs on the network \cite{balduf2022monitoring, balduf2023cloud, cheng2025centralization}. 
A further method proposed to investigate IPFS content is to use gateway logs of gateways that publish IPFS requests \cite{shi2024closer}. 
Although not specifically within the scope of our work, extensive work has also been done on measuring the performance of different operations, such as peer connection or content retrieval \cite{daniel2022passively, lajam2021performance, shi2024closer, cheng2025centralization, trautwein2022ipfs}.

\subsection{Security of IPFS network}
\label{subsec:rel-security}

Research on IPFS security takes mainly two directions. 
The first type of attack is the use of Sybil nodes. 
This attack vector opened up vulnerabilities that allowed the attacker to eclipse individual nodes \cite{prunster2020eclipse} or specific content \cite{netto2025sybil, cholez2024sybil}. 
Although mitigation strategies have been implemented, Sybil node attacks remain a significant challenge in any DHT-based system, such as IPFS. 
Similar to other decentralized systems, such as blockchains, a substantial concern in IPFS is the centralization of key components, which undermines many of the security benefits that decentralization provides. 
Different works investigated the centralization of IPFS and its components \cite{balduf2023cloud, cheng2025centralization, shi2024closer, wei2024eternal}. 
Interestingly, the work of Cheng et al. \cite{cheng2025centralization} reveals a steady shift toward cloud nodes in recent years in the IPFS network since its launch, and today more than 75\% of all nodes are hosted by a few cloud providers. 
Centralized components, such as API gateways, IPNIs, Hydra nodes, and bootstrap nodes, contribute to the centralization of the network. 
As suggested in these works and as we demonstrate in our own work, this centralization has profound implications for protecting the network against censorship attacks.

\subsection{Routing attacks on decentralized systems}
\label{subsec:rel-routing-attacks}

There has been extensive research on BGP routing attacks for other decentralized peer-to-peer systems. 
BGP routing-based attacks, for example, have been shown to enable an attacker to partition networks of different cryptocurrencies \cite{apostolaki2017hijacking, saad2023three}. 
In another study, BGP hijacking attacks were used against crypto-mining pools that utilized the Stratum protocol, enabling the interception of connections between miners and the pool server to waste the computational resources of the miners. 
Although the attack has a different goal, its attack principle of denying a requester from reaching its provider is very close to the attack that we propose in our work~\cite{tran2024routing}. 
\section{Conclusion}
In our work, we demonstrate that due to the centralization of IPFS content providers in a small set of ASes and IP prefixes often operated by large cloud providers and a lack of BGP hijacking protection, a significant portion of content in IPFS is vulnerable to BGP routing-based interception attacks. 
This allows AS-level adversaries to make content and services based on it unavailable to requesters, even with a small set of hijacked IPv4 and IPv6 prefixes of around 60 key prefixes. 
We further presented and evaluated countermeasures involving the proposal for a global collaborative IPFS cluster embedded into the IPFS protocol, ensuring a better decentralization of content within the IPFS network, as well as adding strategic provider nodes to the network that are well protected against BGP hijacking by utilizing RPKI-protected IPv4 and IPv6 prefixes that are announced according to the max-prefix length. 
Together, these two countermeasures promise to make the weak link of provider centralization in IPFS strong enough to stand up against possible BGP routing and DDoS-based censorship attacks.


\cleardoublepage
\bibliographystyle{plain}
\bibliography{main}

\begin{thebibliography}{10}

\bibitem{apostolaki2017hijacking}
Maria Apostolaki, Aviv Zohar, and Laurent Vanbever.
\newblock Hijacking bitcoin: Routing attacks on cryptocurrencies.
\newblock In {\em Proc. IEEE S\&P}, 2017.

\bibitem{pyasn162}
Hadi Asghari and Arman Noroozian.
\newblock pyasn: Offline ip to asn lookup tool.
\newblock \url{https://pypi.org/project/pyasn/}, 2023.
\newblock Python module for fast offline and historical IP-to-ASN lookups. Developed during PhD research on cybersecurity measurements.

\bibitem{balduf2022monitoring}
Leonhard Balduf, Sebastian Henningsen, Martin Florian, Sebastian Rust, and Bj{\"o}rn Scheuermann.
\newblock Monitoring data requests in decentralized data storage systems: A case study of ipfs.
\newblock In {\em Proceedings of the IEEE 42nd International Conference on Distributed Computing Systems (ICDCS)}, 2022.
\newblock Accepted at ICDCS 2022.

\bibitem{balduf2023cloud}
Leonhard Balduf, Navin~V. Keizer, Maciej Korczyński, George Pavlou, Michał Król, Onur Ascigil, and Björn Scheuermann.
\newblock The cloud strikes back: Investigating the decentralization of ipfs.
\newblock \url{https://arxiv.org/abs/2309.16203}, 2023.
\newblock arXiv:2309.16203 [cs.NI].

\bibitem{benet2014ipfs}
Juan Benet.
\newblock {IPFS} - content addressed, versioned, p2p file system.
\newblock \url{https://arxiv.org/abs/1407.3561}, 2014.
\newblock arXiv:1407.3561 [cs.NI], Draft 3.

\bibitem{birge2018bamboozling}
Henry Birge-Lee, Yixin Sun, Anne Edmundson, Jennifer Rexford, and Prateek Mittal.
\newblock {Bamboozling certificate authorities with BGP}.
\newblock In {\em Proc. USENIX Security}, 2018.

\bibitem{irrexplorer2025}
DashCare BV and Stichting NLNOG.
\newblock Irr explorer.
\newblock \url{https://irrexplorer.nlnog.net}, 2025.
\newblock Developed to visualize routing, IRR, and RPKI status for network resources.

\bibitem{caida_asrank}
{CAIDA}.
\newblock As rank: Ranking autonomous systems.
\newblock \url{https://asrank.caida.org/}, 2023.
\newblock Accessed: [Insert Date of Access].

\bibitem{as-relationship}
CAIDA.
\newblock {AS Relationships Dataset}.
\newblock \url{https://www.caida.org/catalog/datasets/as-relationships/}, 2025.

\bibitem{cheng2025centralization}
Ruizhi Cheng, Yuetong Wu, Ashish Kundu, Hugo Latapie, Myungjin Lee, Songqing Chen, and Bo~Han.
\newblock Centralization in the decentralized web: Challenges and opportunities in ipfs data management.
\newblock {\em Proceedings of the ACM Web Conference (WWW)}, April 2025.

\bibitem{cholez2024sybil}
Thibault Cholez and Claudia-Lavinia Ignat.
\newblock Sybil attack strikes again: Denying content access in ipfs with a single computer.
\newblock In {\em Proceedings of the 19th International Conference on Availability, Reliability and Security (ARES)}, pages 1--7. ACM, July 2024.

\bibitem{coldewey2020cloudflare}
Devin Coldewey.
\newblock Cloudflare dns goes down, taking a large piece of the internet with it.
\newblock \url{https://techcrunch.com/2020/07/17/cloudflare-dns-goes-down-taking-a-large-piece-of-the-internet-with-it/}, July 2020.
\newblock Accessed: 2025-07-09.

\bibitem{daniel2022passively}
Erik Daniel and Florian Tschorsch.
\newblock Passively measuring ipfs churn and network size.
\newblock In {\em 2022 IEEE 42nd International Conference on Distributed Computing Systems Workshops (ICDCSW)}, pages 60--65. IEEE, 2022.

\bibitem{de2021accelerating}
Alfonso De~la Rocha, David Dias, and Yiannis Psaras.
\newblock Accelerating content routing with bitswap: A multi-path file transfer protocol in ipfs and filecoin.
\newblock {\em San Francisco, CA, USA (2021)}, 2021.

\bibitem{clouflare-block-ipfs}
Ernesto~Van der Sar.
\newblock {Cloudflare Disables Access to ‘Pirated’ Content on its IPFS Gateway}.
\newblock \url{https://torrentfreak.com/cloudflare-disables-access-to-pirated-content-on-its-ipfs-gateway-230324/}, 2023.

\bibitem{doumanidis2025routing}
Constantine Doumanidis and Maria Apostolaki.
\newblock Routing attacks in ethereum pos: A systematic exploration.
\newblock \url{https://arxiv.org/abs/2505.07713}, 2025.
\newblock arXiv:2505.07713 [cs.NI].

\bibitem{fan2021conman}
Wenjun Fan, Sang-Yoon Chang, Xiaobo Zhou, and Shouhuai Xu.
\newblock Conman: A connection manipulation-based attack against bitcoin networking.
\newblock In {\em Proceedings of the IEEE Conference on Communications and Network Security (CNS)}, pages 101--109. IEEE, 2021.

\bibitem{filebase}
FileBase.
\newblock {The InterPlanetary Development Platform}.
\newblock \url{https://filebase.com/}, 2025.

\bibitem{caida_as_relationships}
CAIDA~Center for Applied Internet Data~Analysis.
\newblock As relationships dataset.
\newblock \url{https://www.caida.org/catalog/datasets/as-relationships/}, 2020.
\newblock Accurate knowledge of AS business relationships is relevant to both technical and economic aspects of the Internet's inter-domain structure. Dataset provided by CAIDA at UC San Diego.

\bibitem{caida_ixps_dataset}
CAIDA~Center for Applied Internet Data~Analysis.
\newblock Internet exchange points (ixps) dataset.
\newblock \url{https://www.caida.org/catalog/datasets/ixps/}, 2020.
\newblock Dataset combining information from PeeringDB, Hurricane Electric, Packet Clearing House, and GeoNames to provide geographic and membership data on IXPs.

\bibitem{gao2001stable}
Lixin Gao and Jennifer Rexford.
\newblock Stable internet routing without global coordination.
\newblock {\em IEEE/ACM TON}, 2001.

\bibitem{goldberg2010securebgp}
Sharon Goldberg and Michael Schapira.
\newblock How secure are secure interdomain routing protocols?
\newblock Technical report, Microsoft Research and Yale University, 2010.
\newblock Full version dated February 23, 2010. Analyzes the effectiveness of secure BGP variants against traffic attraction attacks using empirical AS-level topology simulations.

\bibitem{goldberg2010secure}
Sharon Goldberg, Michael Schapira, Peter Hummon, and Jennifer Rexford.
\newblock How secure are secure interdomain routing protocols.
\newblock {\em ACM SIGCOMM CCR}, 2010.

\bibitem{henningsen2020mapping}
Sebastian Henningsen, Martin Florian, Sebastian Rust, and Björn Scheuermann.
\newblock Mapping the interplanetary filesystem.
\newblock \url{https://arxiv.org/abs/2002.07747}, 2020.
\newblock arXiv:2002.07747 [cs.NI].

\bibitem{crawling}
Sebastian~A. Henningsen, Sebastian Rust, Martin Florian, and Björn Scheuermann.
\newblock Demo: Crawling the ipfs network.
\newblock In {\em Proceedings of the IFIP Networking Conference}, pages 679--680. IEEE, 2020.

\bibitem{karlin2006pretty}
Josh Karlin, Stephanie Forrest, and Jennifer Rexford.
\newblock Pretty good bgp: improving bgp by cautiously adopting routes.
\newblock In {\em Proc. IEEE International Conference on Network Protocols}, 2006.

\bibitem{ipfs_origins_2013}
Protocol Labs.
\newblock Ipfs origins and a new p2p summer (2013–2017).
\newblock \url{https://docs.ipfs.tech/project/history/#ipfs-origins-and-a-new-p2p-summer-2013-2017}, 2023.
\newblock Accessed: 2025-07-08.

\bibitem{ipfs-cluster}
Protocol Labs and Contributors.
\newblock {IPFS Cluster}.
\newblock \url{https://ipfscluster.io/}, 2023.
\newblock Accessed: [Insert Today's Date].

\bibitem{lajam2021performance}
Omar~Abdullah Lajam and Tarek~Ahmed Helmy.
\newblock Performance evaluation of ipfs in private networks.
\newblock In {\em Proceedings of the 2021 4th International Conference on Data Storage and Data Engineering (DSDE)}, pages 77--84. Association for Computing Machinery, 2021.

\bibitem{libp2p}
libp2p.
\newblock {libp2p}.
\newblock \url{https://docs.libp2p.io}, 2025.

\bibitem{libp2p-tls}
libp2p.
\newblock {TLS}.
\newblock \url{https://docs.libp2p.io/concepts/secure-comm/tls/}, 2025.

\bibitem{rpki}
Doug Madory.
\newblock {RPKI ROV Deployment Reaches Major Milestone}.
\newblock \url{https://manrs.org/2024/05/rpki-rov-deployment-reaches-major-milestone/}, 2024.

\bibitem{spain-block-ipfs}
Jeremy Malcolm.
\newblock {No Justification for Spanish Internet Censorship During Catalonian Referendum}.
\newblock \url{https://www.eff.org/deeplinks/2017/10/no-justification-spanish-internet-censorship-during-catalonian-referendum}, 2017.

\bibitem{maymounkov2002kademlia}
Petar Maymounkov and David Mazi{\`{e}}res.
\newblock Kademlia: A peer-to-peer information system based on the {XOR} metric.
\newblock In {\em Revised Papers from the First International Workshop on Peer-to-Peer Systems}, IPTPS '01, pages 53--65, Berlin, Heidelberg, 2002. Springer-Verlag.

\bibitem{multihash}
Multiformats.
\newblock {Multihash}.
\newblock \url{https://multiformats.io/multihash/}, 2025.

\bibitem{nakibly2016website}
Gabi Nakibly, Jaime Schcolnik, and Yossi Rubin.
\newblock $\{$Website-Targeted$\}$ false content injection by network operators.
\newblock In {\em Proc. USENIX Security}, 2016.

\bibitem{netto2025sybil}
Victor Henrique De~Moura Netto, Thibault Cholez, and Claudia-Lavinia Ignat.
\newblock Active sybil attack and efficient defense strategy in ipfs dht.
\newblock {\em arXiv preprint arXiv:2505.01139}, May 2025.

\bibitem{pinata}
Pinata.
\newblock {Crypto's File Storage}.
\newblock \url{https://pinata.cloud/}, 2025.

\bibitem{prunster2020eclipse}
Bernd Prünster, Alexander Marsalek, and Thomas Zefferer.
\newblock Total eclipse of the heart -- disrupting the interplanetary file system.
\newblock \url{https://arxiv.org/abs/2011.00874}, 2020.
\newblock arXiv:2011.00874 [cs.CR].

\bibitem{raman2020worldwide}
Ram~Sundara Raman, Prerana Shenoy, Katharina Kohls, and Roya Ensafi.
\newblock A worldwide view of network-level paths to censorship.
\newblock In {\em Proceedings of the 29th {USENIX} Security Symposium}, pages~--. {USENIX} Association, August 2020.

\bibitem{rekhter2006rfc}
Yakov Rekhter, Tony Li, and Susan Hares.
\newblock Rfc 4271: A border gateway protocol 4 (bgp-4), 2006.

\bibitem{saad2023three}
Muhammad Saad and David Mohaisen.
\newblock Three birds with one stone: Efficient partitioning attacks on interdependent cryptocurrency networks.
\newblock In {\em Proc. IEEE S\&P}, 2023.

\bibitem{sermpezis2018artemis}
Pavlos Sermpezis, Vasileios Kotronis, Petros Gigis, Xenofontas Dimitropoulos, Danilo Cicalese, Alistair King, and Alberto Dainotti.
\newblock Artemis: Neutralizing bgp hijacking within a minute.
\newblock {\em IEEE/ACM TON}, 2018.

\bibitem{ipfs-node-cost}
Pinata Service.
\newblock {How much does an IPFS Pinning Service Cost?}
\newblock \url{https://pinata.cloud/blog/how-much-does-an-ipfs-pinning-service-cost/}, 2023.

\bibitem{shi2024closer}
Ruizhe Shi, Ruizhi Cheng, Bo~Han, Yue Cheng, and Songqing Chen.
\newblock A closer look into ipfs: Accessibility, content, and performance.
\newblock {\em Proceedings of the ACM on Measurement and Analysis of Computing Systems}, 8(2):Article 20, 1--31, June 2024.

\bibitem{siddiqui2018route53}
Aftab Siddiqui.
\newblock What happened? the amazon route 53 bgp hijack to take over ethereum cryptocurrency wallets.
\newblock \url{https://www.internetsociety.org/blog/2018/04/amazons-route-53-bgp-hijack/}, April 2018.
\newblock Accessed: 2025-07-09.

\bibitem{sommese2022ddos}
Raffaele Sommese, KC~Claffy, Roland van Rijswijk-Deij, Arnab Chattopadhyay, Alberto Dainotti, Anna Sperotto, and Mattijs Jonker.
\newblock Investigating the impact of ddos attacks on dns infrastructure.
\newblock In {\em Proceedings of the ACM Internet Measurement Conference (IMC)}, pages 20:1--20:15, Nice, France, 2022. ACM.

\bibitem{sridhar2024censorship}
Srivatsan Sridhar, Onur Ascigil, Navin Keizer, François Genon, Sébastien Pierre, Yiannis Psaras, Etienne Rivière, and Michał Król.
\newblock Content censorship in the interplanetary file system.
\newblock In {\em Network and Distributed System Security (NDSS) Symposium}, San Diego, CA, USA, 2024. The Internet Society.
\newblock arXiv:2307.12212 [cs.CR].

\bibitem{sun2015raptor}
Yixin Sun, Anne Edmundson, Laurent Vanbever, Oscar Li, Jennifer Rexford, Mung Chiang, and Prateek Mittal.
\newblock {RAPTOR: Routing attacks on privacy in Tor}.
\newblock In {\em Proc. USENIX Security}, 2015.

\bibitem{setup-as}
Daryll Swer.
\newblock {How I set up my own Autonomous System}.
\newblock \url{https://blog.apnic.net/2022/07/01/how-i-set-up-my-own-autonomous-system/}, 2022.

\bibitem{unit42-ipfs-malware}
Amanda Tanner, Kristopher Bleich, Anthony Galiette, and Joseph Opacki.
\newblock {Threat Actors Rapidly Adopt Web3 IPFS Technology}.
\newblock \url{https://unit42.paloaltonetworks.com/ipfs-used-maliciously/}, 2023.

\bibitem{cidr}
Geoff~Huston Tony~Bates, Philip~Smith.
\newblock {CIDR Report}.
\newblock \url{https://www.cidr-report.org/as2.0/}, 2025.

\bibitem{tran2024routing}
Muoi Tran, Theo von Arx, and Laurent Vanbever.
\newblock Routing attacks on cryptocurrency mining pools.
\newblock In {\em Proc. IEEE S\&P}, 2024.

\bibitem{trautwein2022ipfs}
Dennis Trautwein, Aravindh Raman, Gareth Tyson, Ignacio Castro, Will Scott, Moritz Schubotz, Bela Gipp, and Yiannis Psaras.
\newblock Design and evaluation of ipfs: A storage layer for the decentralized web.
\newblock In {\em Proceedings of the ACM SIGCOMM 2022 Conference}. ACM, August 2022.

\bibitem{vonarx2023revelio}
Theo von Arx, Muoi Tran, and Laurent Vanbever.
\newblock Revelio: A network-level privacy attack in the lightning network.
\newblock In {\em Proceedings of the 8th IEEE European Symposium on Security and Privacy (EuroS\&P)}, Delft, Netherlands, July 2023. IEEE.

\bibitem{wei2024eternal}
Yiluo Wei, Dennis Trautwein, Yiannis Psaras, Ignacio Castro, Will Scott, Aravindh Raman, and Gareth Tyson.
\newblock The eternal tussle: exploring the role of centralization in $\{$IPFS$\}$.
\newblock In {\em USENIX NSDI}, 2024.

\bibitem{wu2025wall}
Mingshi Wu, Ali Zohaib, Zakir Durumeric, Amir Houmansadr, and Eric Wustrow.
\newblock A wall behind a wall: Emerging regional censorship in china.
\newblock In {\em Proc. IEEE S\&P}, 2025.

\end{thebibliography}

\subsection{Countermeasures - Collaborative IPFS cluster ASes and Prefixes}
\label{appendix:countermeasures-collab-as-prefixes}

\paragraph{Project Gutenberg (Spanish)}
\textbf{ASes}: 54825, 12322, 140527, 14618, 20473, 21928, 24940, 3320, 36352, 396982, 48314, 51167,  56309, 6400, 852\\
\textbf{IPv4 Prefixes}: 114.96.64.0/19, 116.203.0.0/16, 139.178.88.0/22, 142.132.128.0/17, 147.75.63.0/24, 147.75.80.0/22, 147.75.86.0/23, 172.32.0.0/11, 18.232.0.0/14, 185.207.250.0/24, 185.245.99.0/24, 190.80.128.0/17, 203.159.92.0/22, 207.246.80.0/20, 217.224.0.0/11, 217.76.48.0/20, 23.94.131.0/24, 35.201.240.0/20, 45.77.112.0/21, 54.221.0.0/16, 82.64.0.0/14, 88.99.0.0/16, 99.199.0.0/17 \\
\textbf{IPv6 Prefixes}: 2001:41d0::/32, 2001:569:8000::/34, 2604:1380:45e0::/44, 2604:1380:45f0::/44, 2604:1380:4600::/44, 2605:aa80:c000::/34, 2a01:e00::/26

\paragraph{Wikipedia}
\textbf{ASes}: 54825\\
\textbf{IPv4 Prefixes}: 139.178.88.0/22, 147.75.63.0/24, 147.75.80.0/22, 147.75.86.0/23 \\
\textbf{IPv6 Prefixes}: 2604:1380:45e0::/44, 2604:1380:45f0::/44, 2604:1380:4600::/44",
\paragraph{IPFS Websites}
\textbf{ASes}: 54825, 396004\\
\textbf{IPv4 Prefixes}: 139.178.88.0/22, 147.75.63.0/24, 147.75.80.0/22, 147.75.86.0/23, 69.195.151.0/24 \\
\textbf{IPv6 Prefixes}: 2602::/24, 2604:1380:45e0::/44, 2604:1380:45f0::/44, 2604:1380:4600::/44

\subsection{Top 100 ranked ASes - attacker ASes}
\label{appendix:top-rank-ases}
\begin{table}[h!]
    \centering
    \label{table:asns}
    \begin{tabular}{rrl}
        \toprule
        \textbf{Rank} & \textbf{AS} & \textbf{Organization} \\
        \midrule
        1 & 3356 & Level 3 Parent, LLC \\
        2 & 174 & Cogent Communications \\
        3 & 1299 & Arelion Sweden AB \\
        4 & 3257 & GTT Communications Inc. \\
        5 & 2914 & NTT America, Inc. \\
        6 & 6939 & Hurricane Electric LLC \\
        7 & 6453 & TATA COMMUNICATIONS (AMERICA) INC \\
        8 & 6762 & Telecom Italia S.p.A. \\
        9 & 6461 & Zayo Bandwidth \\
        10 & 3491 & PCCW Global, Inc. \\
        11 & 9002 & RETN Limited \\
        12 & 5511 & Orange S.A. \\
        13 & 1273 & Vodafone Group PLC \\
        14 & 12956 & TELEFONICA GLOBAL SOLUTIONS SL \\
        15 & 4637 & Telstra International Limited \\
        16 & 7473 & Singapore Telecommunications \\
        17 & 22356 & Durand do Brasil Ltda \\
        18 & 12389 & PJSC Rostelecom \\
        19 & 3320 & Deutsche Telekom AG \\
        20 & 13786 & Seabras 1 USA, LLC \\
        21 & 9498 & Bharti Airtel Limited \\
        22 & 7195 & EDGEUNO S.A.S \\
        23 & 3216 & PJSC "Vimpelcom" \\
        24 & 37468 & Angola Cables \\
        25 & 16735 & ALGAR TELECOM S/A \\
        26 & 20764 & CJSC RASCOM \\
        27 & 1221 & Telkom SA Ltd. \\
        28 & 2635 & EWE TEL GmbH \\
        29 & 2856 & Telefonica del Sur S.A. \\
        30 & 6830 & Liberty Global Operations B.V. \\
        31 & 394536 & OVH SAS \\
        32 & 4134 & China Telecom \\
        33 & 8867 & VNPT Corp \\
        34 & 35616 & AT\&T Services, Inc. \\
        35 & 9318 & Telecomunicaciones y Servicios S.A. \\
        36 & 3302 & SWISSCOM-AS \\
        37 & 25152 & Korea Telecom \\
        38 & 3786 & TELEFONICA MOVILES ESPANA S.A.U \\
        39 & 16509 & Amazon.com, Inc. \\
        40 & 4837 & China Unicom \\
        41 & 15133 & Vocus Pty Ltd \\
        42 & 4761 & KDDI CORPORATION \\
        43 & 9260 & TATA COMMUNICATIONS (AMERICA) INC \\
        44 & 4812 & Korea Telecom \\
        45 & 5580 & PCCW Global, Inc. \\
        46 & 2828 & XO Communications, LLC \\
        47 & 4809 & China Unicom \\
        48 & 3277 & Zomro B.V. \\
        49 & 13180 & Telecom Italia Sparkle S.p.A. \\
        50 & 58453 & Amazon.com, Inc. \\
    \bottomrule
    \end{tabular}
\end{table}

\begin{table}[h!]
    \centering
    \label{table:asns}
    \begin{tabular}{rrl}
        \toprule
        \textbf{Rank} & \textbf{AS} & \textbf{Organization} \\
        \midrule
        51 & 4755 & NTT DOCOMO, INC. \\
        52 & 4777 & KDDI CORPORATION \\
        53 & 7922 & Comcast Cable Communications, LLC \\
        54 & 10026 & EWE TEL GmbH \\
        55 & 47808 & Vultr Holdings LLC \\
        56 & 4844 & China Mobile Communications Group Co., Ltd. \\
        57 & 34927 & TELEFONICA DEL SUR S.A. \\
        58 & 4766 & KDDI CORPORATION \\
        59 & 38241 & Amazon.com, Inc. \\
        60 & 4758 & NTT DOCOMO, INC. \\
        61 & 4800 & China Unicom \\
        62 & 2497 & L3 Network, LLC \\
        63 & 3333 & T-Mobile USA, Inc. \\
        64 & 48163 & Vultr Holdings LLC \\
        65 & 4768 & KDDI CORPORATION \\
        66 & 16276 & OVH SAS \\
        67 & 35839 & Vultr Holdings LLC \\
        68 & 4739 & Taiwan Fixed Network CO., Ltd \\
        69 & 7401 & Spark New Zealand Ltd \\
        70 & 28867 & Vultr Holdings LLC \\
        71 & 4826 & China Mobile Communications Group Co., Ltd. \\
        72 & 3549 & Level 3 Parent, LLC \\
        73 & 1221 & Telkom SA Ltd. \\
        74 & 2635 & EWE TEL GmbH \\
        75 & 2856 & Telefonica del Sur S.A. \\
        76 & 6830 & Liberty Global Operations B.V. \\
        77 & 394536 & OVH SAS \\
        78 & 4134 & China Telecom \\
        79 & 8867 & VNPT Corp \\
        80 & 35616 & AT\&T Services, Inc. \\
        81 & 9318 & Telecomunicaciones y Servicios S.A. \\
        82 & 3302 & SWISSCOM-AS \\
        83 & 25152 & Korea Telecom \\
        84 & 3786 & TELEFONICA MOVILES ESPANA S.A.U \\
        85 & 16509 & Amazon.com, Inc. \\
        86 & 4837 & China Unicom \\
        87 & 15133 & Vocus Pty Ltd \\
        88 & 4761 & KDDI CORPORATION \\
        89 & 9260 & TATA COMMUNICATIONS (AMERICA) INC \\
        90 & 4812 & Korea Telecom \\
        91 & 5580 & PCCW Global, Inc. \\
        92 & 2828 & XO Communications, LLC \\
        93 & 4809 & China Unicom \\
        94 & 3277 & Zomro B.V. \\
        95 & 13180 & Telecom Italia Sparkle S.p.A. \\
        96 & 58453 & Amazon.com, Inc. \\
        97 & 20473 & The Constant Company, LLC \\
        98 & 3549 & Level 3 Parent, LLC \\
        99 & 209 & CenturyLink Communications, LLC \\
        100 & 4515 & Telstra International Limited \\
        \bottomrule
    \end{tabular}
\end{table}

\end{document}